\documentclass[12pt,journal,onecolumn,draftclsnofoot]{IEEEtran}
\ifCLASSINFOpdf
\else
\fi

\usepackage{graphicx}
\usepackage{amsmath}
\usepackage{cite}
\usepackage{amssymb}
\usepackage{multirow}
\usepackage{color}
\usepackage{epstopdf}
\usepackage{float}

\usepackage{algorithm}
\usepackage{algorithmic}
\usepackage{balance}

\usepackage{flushend}
\usepackage{algorithm}
\usepackage{algorithmic}
\usepackage{tabularx}
\usepackage{array}
\usepackage{booktabs}
\usepackage{makecell}
\usepackage{subfigure}

\newtheorem{remark}{Remark}

\begin{document}

%
\title{Pinching-Antenna-Assisted Index Modulation: Channel Modeling, Transceiver Design, and  Performance Analysis
}

\author{Shuaixin~Yang,~\IEEEmembership{}
	    Yijia~Li,~\IEEEmembership{}
		Yue~Xiao,~\IEEEmembership{}
		Yong~Liang~Guan,~\IEEEmembership{}
		Xianfu~Lei,~\IEEEmembership{}
		and Zhiguo~Ding~\IEEEmembership{}
        \thanks{
	S. Yang, Y. Li and Y. Xiao are with the National Key Laboratory of Wireless Communications, University of Electronic Science and Technology of China, Chengdu 611731, China (e-mail: shuaixin.yang@foxmail.com, nikoeric@foxmail.com, xiaoyue@uestc.edu.cn).

	Y. Guan is with the School of Electrical and Electronic Engineering, Nanyang Technological University, Singapore (e-mail: EYLGuan@ntu.edu.sg).
	
	X. Lei is with the School of Information Science and Technology, Southwest Jiaotong University, Chengdu 610031, China (e-mail: xflei@swjtu.edu.cn).
	
	Z. Ding is with the Department of Computer and Information Engineering, Khalifa University, Abu Dhabi, UAE (e-mail:zhiguo.ding@ieee.org).
}
}

\maketitle

\begin{abstract}
	In this paper, a novel pinching-antenna assisted index modulation (PA-IM) scheme is proposed for improving the spectral efficiency without increasing the hardware complexity, where the information bits are conveyed not only by the conventional M-ary quadrature amplitude modulation (QAM) symbols but also by the indices of pinching antenna (PA) position patterns.
	To realize the full potential of this scheme, this paper focuses on the comprehensive transceiver design, addressing key challenges in signal detection at the receiver and performance optimization at thetransmitter.
	First, a comprehensive channel model is formulated for this architecture, which sophisticatedly integrates the deterministic in-waveguide propagation effects with the stochastic nature of wireless channels, including both large-scale path loss and small-scale fading. Next, to overcome the prohibitive complexity of optimal maximum likelihood (ML) detection, a low-complexity box-optimized sphere decoding (BO-SD) algorithm is designed, which adaptively prunes the search space whilst preserving optimal ML performance. Furthermore, an analytical upper bound on the bit error rate (BER) is derived and validated by the simulations. Moreover, a new transmit precoding method is designed using manifold optimization, which minimizes the BER by jointly optimizing the complex-valued precoding coefficients across the waveguides for the sake of maximizing the minimum Euclidean distance of all received signal points. Finally, the simulation results demonstrate that the proposed PA-IM scheme attains a significant performance gain over its conventional counterparts and that the overall BER of the pinching-antenna system is substantially improved by the proposed precoding design.
\end{abstract}


\IEEEpeerreviewmaketitle

\section{Introduction}         
\IEEEPARstart{T}{he} seminal work of Shannon on channel capacity \cite{Shannon} has established the ultimate performance benchmarks for wireless communications and has profoundly shaped our understanding of the underlying transmission model. However, for decades, the wireless channel, characterized by environmental factors, such as path loss, blockage, and scattering, has been regarded as a stochastic entity that is largely beyond our control. Consequently, the evolution of wireless communication systems, including the sophisticated family of multiple-input multiple-output (MIMO) architectures \cite{mMIMO}, has predominantly focused on designing advanced transceivers capable of mitigating to the deleterious effects of this uncontrollable channel in order to maximize the achievable throughput.

Nevertheless, despite the tremendous success of this adaptive design philosophy, the relentless quest for ubiquitous connectivity and ultra-high data rates toward future wireless communications \cite{6G1,6G2,6G3} is exposing the fundamental limitations of this existing design philosophy. Specifically, the performance gains attainable by solely enhancing transceiver complexity are subject to the law of diminishing returns. More specifically, the hostile propagation environment itself often constitutes the fundamental performance bottleneck. Indeed, in a plethora of practical scenarios, even with sophisticated signal processing at the transmitter and receiver, it's impossible to circumvent a severe blockage or create a favorable propagation path where none inherently exists.
This pivotal realization has heralded a significant paradigm shift in the philosophy of wireless system design: a transition from passively adapting to the channel to actively reconfiguring it \cite{reconfigure1}.

Albeit a concept of recent prominence, the ambition of actively engineering the wireless link has a notable history \cite{reconfigure2,reconfigure3,reconfigure4}. Pioneering efforts in this direction were initially centered on reconfiguring the transceiver, where antenna selection (AS) emerged as one of the foundational techniques for dynamically activating the most advantageous propagation sub-channels \cite{AS1,AS2,AS3,AS4}. These early investigations unequivocally confirmed the tangible benefits of such spatial control. Subsequently, the field has evolved toward more audacious technologies that directly manipulate the propagation environment. A significant advancement in this domain was heralded by the advent of reconfigurable intelligent surfaces (RISs) \cite{RIS1,RIS2}, which directly reconfigure the radio environment by intelligently steering impinging signals. Concurrently, another research frontier has focused on imbuing antennas with physical freedom, culminating in the development of fluid antennas (FAs)  \cite{FA1} and movable antennas (MAs) \cite{MA}. Building upon this compelling technological trajectory, this work investigates a recently proposed reconfigurable technology, namely pinching antenna (PA) \cite{PA1}, in order to elucidate its potential for efficient channel modification in next-generation wireless systems. In the following, we provide a more detailed overview of these spatially reconfigurable technologies to better motivate our investigation.

\subsection{Spatially Reconfigurable Technologies}

\subsubsection{Antenna Selection}
The evolution of spatially reconfigurable technologies can be traced back to pioneering concepts focused on optimizing the transceiver's effective aperture. One of the most foundational among these is the family of AS techniques, which operate on the principle of dynamically activating only a judiciously chosen subset of the total available antennas, thereby selecting the most advantageous propagation sub-channels while deactivating the rest \cite{AS_new1}. A salient benefit of this approach is the significant reduction in both hardware cost and signal processing complexity, which are intrinsically tied to the number of used radio-frequency (RF) chains \cite{AS_new2}.

Furthermore, upon incorporating a holistic power model that accounts for both the transmit and circuit power, AS has emerged as a veritable cornerstone of energy-efficient system design. This is because activating more antennas does not necessarily guarantee enhanced performance, rendering the dynamic optimization of the number of active antennas a pivotal task \cite{AS_new3}. Indeed, the importance of AS is particularly pronounced in the context of multiuser and large-scale MIMO systems, where it is capable of providing substantial energy savings at a negligible loss in spectral efficiency \cite{AS_new4,AS_new5}.

\subsubsection{RIS}
Taking this reconfigurability concept from the transceiver to the propagation environment itself, a truly transformative approach to channel control is heralded by the advent of RISs \cite{RIS1,RIS2,RIS_new1}. These planar metasurfaces are constituted by a large number of low-cost passive elements, and capable of intelligently steering incident signals by controlling their phase shifts, thereby endowing us with the unprecedented ability to sculpt the wireless propagation environment itself.

In practical terms, this unique capability translates into profound performance enhancements across a plethora of metrics. For instance, an RIS can provide a substantial transmit power gain, where the received signal power has been shown to scale quadratically with the number of reflecting elements \cite{RIS_new2}. In the context of MIMO systems, RISs can be invoked to enhance the channel's rank, thus boosting the achievable multiplexing gains \cite{RIS_new3}, or alternatively, be configured to create multiple interference-free beams in multiuser scenarios \cite{RIS_new4}. Furthermore, from an energy-efficiency perspective, they facilitate dramatic reductions in transmit power \cite{RIS_new5}, while simultaneously bolstering physical layer security by actively managing signal propagation toward both legitimate users and potential eavesdroppers \cite{RIS_new6}.

\subsubsection{Fluid Antennas/Movable Antennas}
Moving beyond the discrete, predefined positions inherent to AS, FA/MA systems operate on the compelling principle of dynamically reconfiguring the position and/or shape of radiating elements within a predefined physical space, thereby harnessing spatial diversity at an unprecedentedly fine resolution \cite{FA1}. This intrinsic flexibility translates into profound performance advantages over their conventional fixed-position counterparts \cite{MA}.

More explicitly, FAs/MAs are capable of achieving extreme diversity gains, which in turn directly enhances energy efficiency and reduces outage probability \cite{FA_new1}. Furthermore, this novel positional degree of freedom has heralded the advent of innovative multiple access schemes, such as fluid antenna multiple access (FAMA). This ingenious scheme can effectively mitigate interference without resorting to complex signal processing techniques, such as requiring transmitter-side channel state information (CSI) or employing successive interference cancellation \cite{FA_new2,FA_new3}. This unique capability facilitates rapid channel hardening with fewer active elements \cite{FA_new4} and, most critically, fundamentally circumvents the traditional trade-off between maximizing the desired signal gain and nulling interference \cite{FA_new5}.

\subsection{Pinching Antenna}
Moving beyond systems that primarily reconfigure the channel against small-scale fading, the PA is a particularly novel and promising technology poised to overcome the more fundamental challenges of large-scale path loss, line-of-sight blockage, cost, and complexity. Inspired by DOCOMO's demonstration that attaching simple dielectric materials such as plastic pinches to a waveguide can induce controllable leaky-wave radiation \cite{PA_new1}, PA technology proposes to realize large-scale reconfigurable arrays via long dielectric waveguides, where the propagation pattern is tuned simply by the location of the pinching elements. This approach offers a paradigm shift in terms of implementation simplicity and cost-effectiveness. Instead of relying on complex phase-shifter networks or sophisticated electromagnetic coupling, PAs reconfigure the channel by simply adding, removing, or relocating these low-cost elements.

The unique advantages of PAs are profound. Its primary strength lies in its flexibility and scalability; large-scale antenna arrays spanning tens or even hundreds of meters can be realized with low complexity, allowing for the dynamic creation and strengthening of line-of-sight (LoS) links on demand \cite{PA1}. This effectively transforms a random, fading wireless channel into a quasi-deterministic, ``near-wired" connection, directly overcoming environmental obstacles and severe path loss. Unlike earlier proposals for reconfigurable arrays, this modification is achieved with significantly low-cost and easily deployable components. The initial study in \cite{PA1} has already investigated the potential of PAs for revolutionizing the design of MIMO communication, with subsequent work examining its achievable array gain \cite{PA_new2} and developing low-complexity algorithms to optimize the locations of the pinching elements \cite{PA_new3,PA_new4,PA_new5,PA_new6}. These foundational studies collectively highlight the superiority of PA systems over conventional fixed-position antenna (FPA) systems, underscoring their great potential as a promising reconfigurable antenna technology for next-generation wireless systems.

\subsection{Motivation and Contributions}
Despite the burgeoning research interest in PA systems, sophisticated code designs that fully exploit its flexible antenna structure remain largely unexplored. In this context, the family of index modulation (IM) techniques emerges as a compelling candidate for PA systems [45]. The fundamental principle of IM is to convey additional information bits via the indices of activated communication resources, which may include subcarriers \cite{IM1}, time slots \cite{IM2}, or — most relevant to this work — antennas, a technique also known as spatial modulation (SM) 
\cite{IM3,IM4,IM5}.

Indeed, the amalgamation of IM with various MIMO architectures has been shown to yield profound benefits. For example, space-time block coded spatial modulation (STBC-SM) was conceived in \cite{SM1} by combining SM with space-time block coding (STBC), which facilitates second-order transmit diversity with a reduced number of RF chains. In a similar vein, the authors of \cite{SM2} intrinsically amalgamated the Vertical Bell Labs Space-Time (V-BLAST) architecture \cite{SM3} with SM, giving rise to SM-VBLAST, which further increases the transmission rate. More recently, the concept of RIS has also been brought into the realm of IM, yielding significant gains in both spectral efficiency and bit error rate (BER) performance, particularly in the low signal-to-noise ratio (SNR) regime \cite{SM4}.

In order to harness the unique synergy between the reconfigurability of PA systems and the spectral efficiency of index modulation (IM), this paper proposes a novel PA-assisted IM (PA-IM) scheme for MIMO communication. Our primary motivation stems from the insight that while IM can theoretically be applied to other reconfigurable technologies like RIS or FAs/MAs, the PA architecture presents a uniquely suitable platform, a choice motivated by several compelling advantages. Unlike RIS, which modulates the phase of reflected signals, PA directly controls the locations of the radiation sources, offering a more direct and potentially lower-loss link. Furthermore, compared to the potential mechanical complexity of FAs/MAs, the pinching mechanism offers a simpler, more scalable, and cost-effective way to create the large number of distinct spatial patterns required for high-rate IM. The main contributions of this paper can be summarized as follows.
\begin{itemize}
	\item A new MIMO transmission scheme, termed PA-IM, is proposed. In the proposed PA-IM framework, the information bits are conveyed not only by the conventional $M$-ary quadrature amplitude modulation (QAM) symbols but also by the indices of PA position patterns.
	\item We formulate a comprehensive channel model for PA-aided communication that integrates the deterministic phase shifts from in-waveguide propagation with the stochastic effects of the wireless channel, including both large-scale path loss and small-scale fading.
	\item We design a box-optimized sphere decoding (BO-SD) algorithm for the proposed PA-IM system, which is capable of adaptively pruning the search space to preserve optimal maximum likelihood (ML) performance while significantly reducing detection complexity.
	\item To further enhance the PA-IM system's performance from the transmitter's perspective, the upper bound on the BER for the proposed PA-IM system is derived and validated by simulation results, which then serves as the basis for designing a new minimum-BER based diagonal transmit precoding  method that jointly optimizes transmit powers and phases to maximize the minimum Euclidean distance of all received signal points.
	\item Simulation results show that the proposed PA-IM scheme can attain significant performance gain over its traditional counterpart, and the overall BER gain of the proposed method is signiﬁcantly improved with the designed transmit precoding method.
\end{itemize}

The remainder of this paper is organized as follows. Section \ref{sec:system_model} describes the PA-IM scheme as well as the associated channel model. In Section \ref{sec:bo_sd}, we propose the BO-SD algorithm for the proposed PA-IM system,  while in Section \ref{sec:performance_analysis}, the BER performance analysis of the PA-IM system is conducted. In Section \ref{sec:precoding_design}, we design a transmit precoding scheme toward BER performance optimization. Simulation results and performance comparisons are presented in Section \ref{sec:simulation}, and finally, concluding remarks are drawn in Section \ref{sec:conclusion}.

\emph{Notation}: Bold lowercase and uppercase letters denote column vectors and matrices, respectively. $\mathbb{R}$ and $\mathbb{C}$ denote the sets of real and complex numbers. The operators $(\cdot)^T$, $(\cdot)^H$, $(\cdot)^*$, $|\cdot|$, $\|\cdot\|_2$, and $\|\cdot\|_F$ represent the transpose, Hermitian transpose, conjugate, determinant, Euclidean norm, and Frobenius norm, respectively. $\odot$ and $\otimes$ are the Hadamard and Kronecker products. $\text{vec}(\cdot)$ and $\nabla$ represent the vectorization and gradient operators. $\text{diag}(\cdot)$ and $\text{blkdiag}(\cdot)$ create a diagonal and a block-diagonal matrix from their arguments. $\mathbb{E}\{\cdot\}$ denotes statistical expectation. $\mathcal{R}\{\cdot\}$ and $\mathcal{I}\{\cdot\}$ extract the real and imaginary parts. $\lfloor \cdot \rfloor$ and $\binom{n}{k}$ represent the floor function and the binomial coefficient. We use $\lfloor x\rfloor_{2^p}$ for the largest integer less than or equal to $x$ that is an integer power of two.
	
\section{Pinching Antenna-Assisted Index  Modulation}\label{sec:system_model}

To lay the groundwork for subsequent design and analysis, we first establish a comprehensive system and channel model that is specific to this novel architecture.

\begin{figure}[htbp]
	\centering
	\includegraphics[width=3.5in]{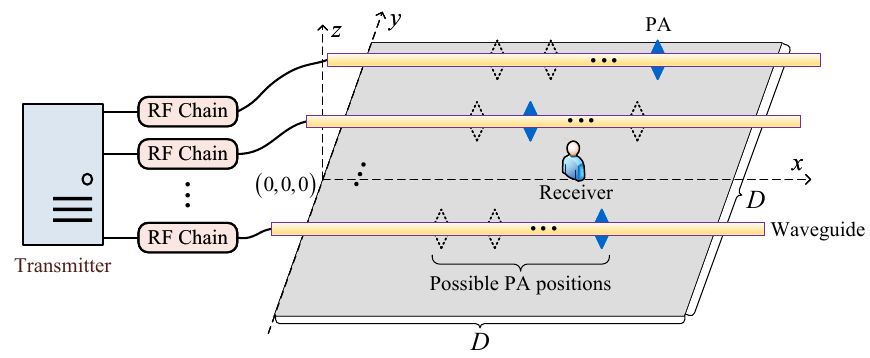}
	\caption{System model of the proposed PA-IM system.}
	\label{fig1}
\end{figure}

Consider the point-to-point PA-IM system depicted in Fig.~\ref{fig1}, which comprises a transmitter employing $N_{\rm wg}$ waveguides and a receiver equipped with $N_{\rm r}$ FPAs, operating within a square area of side length $D$.
Specifically, each of the $N_{\rm wg}$ waveguides is associated with a dedicated RF chain.

The core principle of the PA-IM scheme leverages the practical implementation of PA systems, where elements are preinstalled along waveguides. This allows information to be conveyed not only by amplitude and phase modulation (APM) symbols but also by the indices of dynamically activated positions, rather than physically moving antennas. Specifically, on each waveguide, a signal is radiated from $N_{\rm a}$ out of $N_{\rm t}$ andidate pre-installed PA positions, a method that inherently facilitates index modulation. The details of selecting activated positions based on the information bits will be discussed in Section \ref{sec:signal_model}.

\subsection{Channel Model}\label{sec:channel_model}

In the extant literature concerning PA technology, the predominant channel models rely on the Friis transmission equation, which is typically invoked for characterizing free-space propagation.
While such models offer the advantage of concisely characterizing the signal power decay under LoS conditions and directly relating antenna positional shifts to large-scale fading variations, they generally fail to account for the ubiquitous multipath phenomena prevalent in practical scenarios, particularly within complex urban environments.
Furthermore, these models inherently assume LoS-dominated propagation, which often does not hold in realistic deployments.
More importantly, the significant impact of shadow fading, which results from the random signal obstruction and reflection by obstacles such as buildings and terrain, on the local mean received power has been typically overlooked.
Hence, accurately modeling this shadow fading component is crucial for realistically predicting the system's coverage and performance, especially in intricate propagation environments.

To facilitate the channel modeling, the three-dimensional Cartesian coordinate system is established in Fig.~\ref{fig1}.
Assume that the receiver employs a fixed-position uniform linear array (ULA), where the positions of its $N_{\rm r}$ antenna elements are represented by the position vectors $\{ \mathbf{r}_m \in \mathbb{R}^3 \}_{m=1}^{N_{\rm r}}$, with the coordinates of the $m$-th element being given by $\mathbf{r}_m = [x_{{\rm r},m}, y_{{\rm r},m}, z_{{\rm r},m}]^T$.
As for the transmitter, which is equipped with $N_{\rm wg}$ waveguides and PAs, the position of the $j$-th activated PA on the $n$-th waveguide is denoted by $\mathbf{t}_{n,j}^{\rm wg} = [x_{n,j}^{\rm t,wg}, y_{n,j}^{\rm t,wg}, z_{n,j}^{\rm t,wg}]^T$.
The signals are fed into the waveguides via the RF chains at their feed points, whose positions are denoted by $\mathbf{t}_{n}^{\rm fp} = [x_{n}^{\rm t,fp}, y_{n}^{\rm t,fp}, z_{n}^{\rm t,fp}]^T$.

Against this background, the channel matrix between the $n$-th waveguide and the receiver is denoted by $\mathbf{H}_n$. Its $(i,j)$-th element, which represents the channel coefficient between the $j$-th activated PA on the $n$-th waveguide and the $i$-th antenna of the receiver, can be formulated as
\begin{equation}
	\label{eq:channel_coefficient}
	h_{i,j}^n = \underbrace{\sqrt{\beta_{i,j}^n}}_{\text{Large-scale}} \cdot \underbrace{\left( \sqrt{\frac{K_{i,j}^n}{K_{i,j}^n + 1}} \bar{h}_{i,j}^n + \sqrt{\frac{1}{K_{i,j}^n + 1}} \tilde{h}_{i,j}^n \right)}_{\text{Small-scale}} \cdot \underbrace{\gamma_{i,j}^n}_{\text{Phase-shift}},
\end{equation}
where $\beta_{i,j}^n$ denotes the large-scale channel fading coefficient, encompassing both path loss and shadow fading.
The term in the parentheses characterizes the small-scale fading, where $\bar{h}_{i,j}^n$ represents the deterministic LoS component, while $\tilde{h}_{i,j}^n$ captures the non-line-of-sight (NLoS) component, which is constituted by the multitude of scattered multipath contributions.
Furthermore, the parameter $K_{i,j}^n$ quantifies the power ratio between the LoS and NLoS components, which is commonly referred to as the Rician factor.
It is worth noting that all PAs are allocated along their respective waveguides at a certain distance from their feed points, which inevitably imposes phase shifts on their leaked signals~\cite{PA1}.
To account for this phenomenon, the phase shift term is introduced as $\gamma_{i,j}^n = \exp\left(-\frac{j2\pi}{\lambda_g} \| \mathbf{t}_{n}^{\rm fp} - \mathbf{t}_{n,j}^{\rm wg} \|_2\right)$, where the wavelength in the waveguide is defined as $\lambda_{\rm g} = \lambda / \epsilon_{\text{eff}}$.
Here, $\lambda = c/f_{\rm c}$ is the free-space wavelength at the carrier frequency $f_{\rm c}$, where $c$ is the speed of light in a vacuum, while $\epsilon_{\text{eff}} > 1$ signifies the effective refractive index of the waveguide.

Consequently, the MIMO sub-channel can be expressed in a more compact matrix form as
\begin{equation}
	\label{eq:subchannel_matrix}
	\mathbf{H}_n = \mathbf{B}_n \odot \left( \mathbf{K}_{\rm LoS}^{(n)} \odot \mathbf{\bar{H}}_n + \mathbf{K}_{\rm NLoS}^{(n)} \odot \mathbf{\tilde{H}}_n \right) \odot \mathbf{\Gamma}_n,
\end{equation}
where the matrices $\mathbf{B}_n$, $\mathbf{K}_{\rm LoS}^{(n)}$, $\mathbf{\bar{H}}_n$, $\mathbf{K}_{\rm NLoS}^{(n)}$, $\mathbf{\tilde{H}}_n$, and $\mathbf{\Gamma}_n$ contain the element-wise counterparts of $\sqrt{\beta_{i,j}^n}$, $\sqrt{K_{i,j}^n / (K_{i,j}^n + 1)}$, $\bar{h}_{i,j}^n$, $\sqrt{1 / (K_{i,j}^n + 1)}$, $\tilde{h}_{i,j}^n$, and $\gamma_{i,j}^n$, respectively.

Finally, by aggregating the sub-channels corresponding to all the waveguides, the overall channel matrix between the transmitter and the receiver can be formulated as
\begin{equation}
	\label{eq:overall_channel_matrix}
	\mathbf{H} = \left[ \begin{array}{cccc} \mathbf{H}_1 & \mathbf{H}_2 & \cdots & \mathbf{H}_{N_{\rm wg}} \end{array} \right].
\end{equation}

\subsection{Signal Model}\label{sec:signal_model}
In the proposed PA-IM scheme, the concept of IM is invoked at the transmitter.
Specifically, the incoming bit stream is partitioned into two constituent streams.
The first stream, containing $N_{\rm wg} \log_2 M$ bits, modulates the APM symbols, while the second stream of $N_{\rm wg} \left\lfloor \log_2 \binom{N_{\rm t}}{N_{\rm a}} \right\rfloor$ bits designates the specific set of activated PA positions across all waveguides.

To elaborate, for the $k$-th waveguide, a block of $\log_2 M$ bits is modulated to an $M$-QAM symbol, denoted by $s_k \in \mathcal{S}$, where $\mathcal{S}$ is the $M$-ary constellation set.
Concurrently, a block of $\left\lfloor \log_2 \binom{N_{\rm t}}{N_{\rm a}} \right\rfloor$ bits is used to select a unique combination of $N_{\rm a}$ activated positions from the $N_{\rm t}$ candidates, which is represented by the index set $I_k = \{i_{k,1}, i_{k,2}, \dots, i_{k,N_{\rm a}}\}$, indicating that the $i_{k,m}$-th candidate position is activated for the $m$-th PA.
The resultant transmit signal vector for the $k$-th waveguide, $\mathbf{x}_k \in \mathbb{C}^{N_{\rm t} \times 1}$, can thus be formulated as
\begin{equation}
	\label{eq:waveguide_tx_vector}
	\mathbf{x}_k = \sum_{m=1}^{N_{\rm a}} s_k \mathbf{e}_{i_{k,m}},
\end{equation}
where $\mathbf{e}_{i_{k,m}}$ is the $i_{k,m}$-th column of the identity matrix $\mathbf{I}_{N_{\rm t}}$.
Note that since all $N_{\rm a}$ activated PAs are located on the same waveguide, they radiate the identical baseband symbol $s_k$ \cite{PA1}\footnote{Although phase shifts exist among the leakage signals from different PAs on the same waveguide, this effect is incorporated into the channel model, as detailed in Section \ref{sec:channel_model}, allowing these signals to be treated as identical.}.
By aggregating the contributions from all waveguides, the overall transmit signal vector $\mathbf{x} \in \mathbb{C}^{N_{\rm t} N_{\rm wg} \times 1}$ can be formulated as the concatenation of the individual waveguide vectors, i.e.,
\begin{equation}
	\label{eq:overall_tx_vector_concatenated}
	\mathbf{x} = \left[ \mathbf{x}_1^T, \mathbf{x}_2^T, \dots, \mathbf{x}_{N_{\rm wg}}^T \right]^T.
\end{equation}

For a more compact representation, 
denote the specific activation pattern across all $N_{\rm wg}$ waveguides by $\mathcal{I} = \{I_1, I_2, \dots, I_{N_{\rm wg}}\}$.
This pattern $\mathcal{I}$ determines the structure of a sparse selection matrix, $\mathbf{E}_{\mathcal{I}} \in \{0,1\}^{N_t N_{\rm wg} \times N_a N_{\rm wg}}$, which can be formulated as a block diagonal matrix:
\begin{equation}
	\label{eq:selection_matrix}
	\mathbf{E}_{\mathcal{I}} = \mathrm{blkdiag}\left( \mathbf{E}_{I_1}, \mathbf{E}_{I_2}, \dots, \mathbf{E}_{I_{N_{\rm wg}}} \right),
\end{equation}
where each constituent block $\mathbf{E}_{I_k} = [\mathbf{e}_{i_{k,1}}, \dots, \mathbf{e}_{i_{k,N_{\rm a}}}]$ is constructed by concatenating the basis vectors corresponding to the activated positions in $I_k$.
Furthermore, the vector of non-zero symbols to be transmitted, $\mathbf{x}_{\mathcal{I}} \in \mathbb{C}^{N_{\rm a} N_{\rm wg} \times 1}$, is constructed from the independent baseband symbols $\mathbf{s} = [s_1, \dots, s_{N_{\rm wg}}]^T$ as
\begin{equation}
	\label{eq:compact_tx_vector}
	\mathbf{x}_{\mathcal{I}} = \mathbf{s} \otimes \mathbf{1}_{N_a},
\end{equation}
where ${\mathbf 1}_{N_{\rm a}}$ is an all-one vector with length of $N_{\rm a}$. By harnessing this framework, the overall sparse transmit vector $\mathbf{x}$ of Eq.~(\ref{eq:overall_tx_vector_concatenated}) can be equivalently expressed as
\begin{equation}
	\label{eq:overall_tx_vector_compact}
	\mathbf{x} = \mathbf{E}_{\mathcal{I}} \mathbf{x}_{\mathcal{I}}.
\end{equation}
This formulation explicitly separates the symbol-level information, encapsulated in $\mathbf{x}_{\mathcal{I}}$, from the position-level information, encapsulated in $\mathbf{E}_{\mathcal{I}}$. The set containing all the legitimate transmit signals is denoted by $\mathbb X$.

Finally, the spectral efficiency of the proposed PA-IM system can be readily calculated as
\begin{equation}
	\label{eq:spectral_efficiency}
	\eta = \underbrace{N_{\rm wg} \left\lfloor \log_2 \binom{N_{\rm t}}{N_{\rm a}} \right\rfloor}_{\text{IM Data Rate}} + \underbrace{N_{\rm wg} \log_2 M}_{\text{APM Data Rate}}\quad {\rm bits/s/Hz}.
\end{equation}

Upon transmitting the symbol vector $\mathbf{x}$ over the fading channel, the received signal vector $\mathbf{y} \in \mathbb{C}^{N_{\rm r} \times 1}$ can be formulated as
\begin{equation}
	\label{eq:received_signal}
	\mathbf{y} = \sqrt{\rho} \mathbf{H} \mathbf{x} + \mathbf{n} = \sqrt{\rho} \mathbf{H} \mathbf{E}_{\mathcal{I}} \mathbf{x}_{\mathcal{I}} + \mathbf{n},
\end{equation}
where $\rho = \frac{P_{\rm t}}{N_{\rm wg} N_{\rm a}}$ is the normalized transmit power per activated antenna with the transmit power of $P_{\rm t}$, $\mathbf{H}$ is the overall channel matrix in Eq.~(\ref{eq:overall_channel_matrix}), and $\mathbf{n} \in \mathbb{C}^{N_{\rm r} \times 1}$ is the noise vector, whose entries are assumed to be independent and identically distributed (i.i.d.) complex Gaussian random variables following the distribution $\mathcal{CN}(0, N_0)$.

Consequently, the optimal ML detector for the PA-IM system jointly estimates both the transmitted symbols $\mathbf{s}$ and the activated position pattern $\mathcal{I}$, which can be expressed as
\begin{equation}
	\label{eq:ml_detector}
	(\hat{\mathbf{s}}, \hat{\mathcal{I}}) = \arg \min_{\mathbf{s} \in \mathcal{S}^{N_{\rm wg}}, \mathcal{I} \in \mathbb{I}} \left\| \mathbf{y} - \sqrt{\rho} \mathbf{H} \mathbf{E}_{\mathcal{I}} (\mathbf{s} \otimes \mathbf{1}_{N_{\rm a}}) \right\|_2^2,
\end{equation}
where $\mathbb{I}$ represents the set of all legitimate activation patterns.

\section{Box-Optimized Sphere Decoding Algorithm for the Proposed PA-IM System}\label{sec:bo_sd}
While the conventional ML detection theoretically achieves the optimal BER performance, its computational complexity often escalates to prohibitive levels in practice.
This substantial complexity stems from the requisite exhaustive search across a vast combinatorial space, which encompasses all legitimate combinations of the modulated symbols and of the pertinent antenna indices.

Against this background, we are motivated to conceive a beneficial reduced-complexity detector.
Specifically, we propose a novel BO-SD algorithm, which is tailored for the proposed PA-IM system.
The BO-SD framework is conceived to substantially curtail the computational overhead associated with achieving an optimal ML detection performance, thereby rendering the PA-IM system more viable for practical deployment.

\subsection{Low-Complexity Sphere Decoding}
For a given PA position pattern $\mathcal{I} \in \mathbb{I}$, the system model of Eq.~(\ref{eq:received_signal}) can be expressed as
\begin{equation}
	\label{eq:sys_model_equivalent}
	\mathbf{y} = \sqrt{\rho} \underbrace{\mathbf{H} \mathbf{E}_{\mathcal{I}}  (\mathbf{I}_{N_{\rm wg}} \otimes \mathbf{1}_{N_{\rm a}})}_{\triangleq \mathbf{H}_{\mathcal{I}, \mathrm{eq}}} \mathbf{s} + \mathbf{n},
\end{equation}
where $\mathbf{H}_{\mathcal{I}, \mathrm{eq}} \in \mathbb{C}^{N_r \times N_{\rm wg}}$ is the equivalent channel matrix.

The ML detection problem then reduces to finding the symbol vector $\mathbf{s}$ that minimizes the squared Euclidean distance, which for a specific pattern $\mathcal{I}$ can be formulated as
\begin{equation}
	\label{eq:ml_problem_simplified}
	\hat{\mathbf{s}}_{\mathcal{I},{\rm ML}} = \arg \min_{\mathbf{s} \in \mathcal{S}^{N_{\rm wg}}} \left\| \mathbf{y}' - \mathbf{H}_{\mathcal{I}, \mathrm{eq}} \mathbf{s} \right\|_2^2,
\end{equation}
where the normalized received vector is defined as $\mathbf{y}' \triangleq \mathbf{y}/\sqrt{\rho}$.
To facilitate a low-complexity solution, the classic QR decomposition of the equivalent channel matrix is used to yield $\mathbf{H}_{\mathcal{I}, \mathrm{eq}} = \mathbf{Q}_{\mathcal{I}} \mathbf{R}_{\mathcal{I}}$, where $\mathbf{Q}_{\mathcal{I}}$ is a unitary matrix and $\mathbf{R}_{\mathcal{I}}$ is an upper triangular matrix.
This allows the ML problem of Eq.~(\ref{eq:ml_problem_simplified}) to be transformed into the following equivalent form
\begin{equation}
	\label{eq:ml_problem_qr}
	\hat{\mathbf{s}}_{\mathcal{I}} = \arg \min_{\mathbf{s} \in \mathcal{S}^{N_{\rm wg}}} \left\| \mathbf{z}_{\mathcal{I}} - \mathbf{R}_{\mathcal{I}} \mathbf{s} \right\|_2^2,
\end{equation}
where $\mathbf{z}_{\mathcal{I}} \triangleq \mathbf{Q}_{\mathcal{I}}^H \mathbf{y}'$.

Observe that the cardinality of the search space $\mathcal{S}^{N_{\rm wg}}$ in Eq.~(\ref{eq:ml_problem_qr}) is $M^{N_{\rm wg}}$, which becomes prohibitively large upon increasing the number of waveguides $N_{\rm wg}$ or the modulation order $M$.
Thus, the fundamental principle of sphere decoding (SD) is to circumvent this exhaustive search by only considering candidate vectors that reside within a hypersphere of radius $d$ centered at $\mathbf{z}_{\mathcal{I}}$, i.e.,
\begin{equation}
	\label{eq:sd_constraint}
	\left\| \mathbf{z}_{\mathcal{I}} - \mathbf{R}_{\mathcal{I}} \mathbf{s} \right\|_2^2 \leq d^2.
\end{equation}

\subsubsection{Box-Optimized Radius Initialization}
The selection of an appropriate initial radius $d$ poses a significant challenge.
The optimal radius would be the Euclidean distance associated with the true ML solution, i.e., $d_{\rm ML} = \|\mathbf{z}_{\mathcal{I}} - \mathbf{R}_{\mathcal{I}} \hat{\mathbf{s}}_{{\mathcal I},{\rm ML}}\|_2$.
However, $\hat{\mathbf{s}}_{{\mathcal I},{\rm ML}}$ is unknown a priori.
If the chosen radius is smaller than $d_{\rm ML}$, the search may fail to find any feasible candidates.
Conversely, an excessively large radius may result in a search space that is too vast, leading to a prohibitive computational cost, an issue that is particularly critical in low SNR scenarios.

To address the challenge of radius selection, we utilize a novel box-optimized technique, which is inspired by the principles outlined in~\cite{BO}.
The core idea is to obtain a high-quality estimate of the ML solution by solving a relaxed version of the problem, where the discrete constellation constraint is replaced by a continuous box constraint.
Specifically, the auxiliary box-constrained quadratic program (QP) is established as
\begin{subequations}
	\label{eq:box_optimization}
	\begin{align}
		\label{eq:P1_objective}
		&\left( {{\rm P}1} \right)\quad\min_{\substack{\mathbf s}\in {\mathbb C}^{N_{\rm wg}}}\quad {\left\| {{\bf{z}}_{\mathcal I} - {{\bf{R}}_{\mathcal I}}{\mathbf s}} \right\|_2^2}\\
		&s.t.\quad \min\left(\Re\{\mathcal{S}\}\right)\cdot\mathbf{1}\leq \Re\left({\mathbf s}\right)\leq\max\left(\Re\{\mathcal{S}\}\right)\cdot\mathbf{1},\\
		&\phantom{s.t.\quad}\min\left(\Im\{\mathcal{S}\}\right)\cdot\mathbf{1}\leq \Im\left({\mathbf s}\right)\leq\max\left(\Im\{\mathcal{S}\}\right)\cdot\mathbf{1},
	\end{align}
\end{subequations}
which is a convex optimization problem, and hence can be efficiently solved by those off-shelf optimization solvers or active set algorithms \cite{active_set}.
After obtaining the optimal solution $\tilde{\mathbf{s}}_{\mathcal{I}}$ of (P1), the initial radius can be calculated as $d = \| \mathbf{z}_{\mathcal{I}} - \mathbf{R}_{\mathcal{I}} \tilde{\mathbf{s}}_{\mathcal{I}} \|_2$.

\subsubsection{Layered Detection with Box-Optimized Pruning}
In order to determine the candidate symbol set of $s_{N_{\rm wg}}$, we reformulate \eqref{eq:P1_objective} as
\begin{equation}
	\begin{split}
		&{\left\| {\left[ {\begin{array}{*{20}{c}}
						{{\bf{z}}_1^{\left( {{N_{\rm wg}}} \right)}}\\
						{{\bf{z}}_2^{\left( {{N_{\rm wg}}} \right)}}
				\end{array}} \right]- \left[ {\begin{array}{*{20}{c}}
						{{\bf{R}}_{1,1}^{\left( {{N_{\rm wg}}} \right)}}&{{\bf{R}}_{1,2}^{\left( {{N_{\rm wg}}} \right)}}\\
						{{\bf{0}}_{2,1}^{\left( {{N_{\rm wg}}} \right)}}&{{\bf{R}}_{2,2}^{\left( {{N_{\rm wg}}} \right)}}
				\end{array}} \right]\left[ {\begin{array}{*{20}{c}}
						{{\bf{s}}_1^{\left( {{N_{\rm wg}}} \right)}}\\
						{{s_{{N_{\rm wg}}}}}
				\end{array}} \right]} \right\|_2^2}\\
		=& \underbrace{{\left\| {{\bf{z}}_1^{\left( {{N_{\rm wg}}} \right)} - {\bf{R}}_{1,1}^{\left( {{N_{\rm wg}}} \right)}{\bf{s}}_1^{\left( {{N_{\rm wg}}} \right)} - {\bf{R}}_{1,2}^{\left( {{N_{\rm wg}}} \right)}s_{N_{\rm wg}}} \right\|_2^2}}_{T_1^{\left(N_{\rm wg}\right)}}\\
		&+ {\left\| {{\bf{z}}_2^{\left( {{N_{\rm wg}}} \right)} - {\bf{R}}_{2,2}^{\left( {{N_{\rm wg}}} \right)}s_{N_{\rm wg}}} \right\|_2^2}\\
	\le& {d^2},
	\end{split}
\end{equation}
where ${{\bf{z}}_1^{\left( {{N_{\rm wg}}} \right)}}\in {\mathbb C}^{N_{\rm r}-1}$, ${{\bf{z}}_2^{\left( {{N_{\rm wg}}} \right)}}\in {\mathbb C}$, ${{\bf{R}}_{1,1}^{\left( {{N_{\rm wg}}} \right)}}\in{\mathbb C}^{\left(N_{\rm r}-1\right)\times\left(N_{\rm wg}-1\right)}$, ${{\bf{R}}_{1,2}^{\left( {{N_{\rm wg}}} \right)}}\in{\mathbb C}^{\left(N_{\rm r}-1\right)\times 1}$, ${{\bf{0}}_{2,1}^{\left( {{N_{\rm wg}}} \right)}}\in {\mathbb C}^{1\times\left({N_{\rm wg}-1}\right)}$, ${{\bf{R}}_{2,2}^{\left( {{N_{\rm wg}}} \right)}}\in{\mathbb C}$ and ${{\bf{s}}_1^{\left( {{N_{\rm wg}}} \right)}}\in{\mathcal S}^{N_{\rm wg}-1}$.
The candidate for $s_{N_{\rm wg}}$ can be obtained by determining the symbols satisfying
\begin{equation}
	{\left\| {{\bf{z}}_2^{\left( {{N_{\rm{wg}}}} \right)} - {\bf{R}}_{2,2}^{\left( {{N_{\rm{wg}}}} \right)}{s_{{N_{\rm{wg}}}}}} \right\|_2^2} \le {d^2} - T_1^{\left( {{N_{\rm{wg}}}} \right)} \le {d^2}.
\end{equation}
In order to further obtain a tighter bound and reduce the number of feasible values, we focus on deriving a lower bound on the remaining term $T_1^{\left(N_{\rm wg}\right)}$.

More explicitly, we formulate an auxiliary MIMO detection problem pertaining to the undemodulated symbol vector as
\begin{subequations}
\begin{align}
	&\left( {{\rm P}2\text{-}N_{\rm wg}} \right)\notag\\ &\mathop {\min }\limits_{{\bf{s}}_1^{\left( {{N_{\rm{wg}}}} \right)} \in {{\mathbb C}^{{N_{\rm wg}-1}}}} {\left\| {{\bf{z}}_1^{\left( {{N_{\rm{wg}}}} \right)} - {\bf{R}}_{1,2}^{\left( {{N_{\rm{wg}}}} \right)}{s_{{N_{\rm{wg}}}}} - {\bf{R}}_{1,1}^{\left( {{N_{\rm{wg}}}} \right)}{\bf{s}}_1^{\left( {{N_{\rm{wg}}}} \right)}} \right\|_2^2}\\
	&s.t.\quad \min\left(\Re\{\mathcal{S}\}\right)\cdot\mathbf{1}\leq \Re\left({{\bf{s}}_1^{\left( {{N_{\rm{wg}}}} \right)}}\right)\leq\max\left(\Re\{\mathcal{S}\}\right)\cdot\mathbf{1},\\
	&\phantom{s.t.\quad}\min\left(\Im\{\mathcal{S}\}\right)\cdot\mathbf{1}\leq \Im\left({{\bf{s}}_1^{\left( {{N_{\rm{wg}}}} \right)}}\right)\leq\max\left(\Im\{\mathcal{S}\}\right)\cdot\mathbf{1}.
\end{align}
\end{subequations}
Note that $\left( {{\rm P}2\text{-}N_{\rm wg}} \right)$ is also a convex box-constrained quadratic program, which can be solved efficiently using the same methods as $\left( {{\rm P}1} \right)$. After obtaining the optimal solution ${\bf{\tilde s}}_1^{\left( {{N_{\rm{wg}}}} \right)}$, it's trivial to validate that
\begin{equation}
	\begin{split}
		&{\left\| {{\bf{z}}_1^{\left( {{N_{\rm{wg}}}} \right)} - {\bf{R}}_{1,2}^{\left( {{N_{\rm{wg}}}} \right)}{s_{{N_{\rm{wg}}}}} - {\bf{R}}_{1,1}^{\left( {{N_{\rm{wg}}}} \right)}{\bf{\tilde s}}_1^{\left( {{N_{\rm{wg}}}} \right)}} \right\|_2^2} \\
		\le& {\left\| {{\bf{z}}_1^{\left( {{N_{\rm{wg}}}} \right)} - {\bf{R}}_{1,2}^{\left( {{N_{\rm{wg}}}} \right)}{s_{{N_{\rm{wg}}}}} - {\bf{R}}_{1,1}^{\left( {{N_{\rm{wg}}}} \right)}{\bf{s}}_1^{\left( {{N_{\rm{wg}}}} \right)}} \right\|_2^2},\forall {\bf{s}}_1^{\left( {{N_{\rm{wg}}}} \right)} \in {{\mathcal S} ^{{N_{\rm wg}} - 1}},
	\end{split}
\end{equation}
thus we can obtain the lower bound on $T_1^{\left(N_{\rm wg}\right)}$ as
\begin{equation}
	C_1^{\left( {{N_{\rm{wg}}}} \right)} = {\left\| {{\bf{z}}_1^{\left( {{N_{\rm{wg}}}} \right)} - {\bf{R}}_{1,2}^{\left( {{N_{\rm{wg}}}} \right)}{s_{{N_{\rm{wg}}}}} - {\bf{R}}_{1,1}^{\left( {{N_{\rm{wg}}}} \right)}{\bf{\tilde s}}_1^{\left( {{N_{\rm{wg}}}} \right)}} \right\|_2^2}.
\end{equation}

Then, the scope of the search can be further narrowed by
\begin{equation}\label{eq:scope_s_nwg}
	{\left\| {{\bf{z}}_2^{\left( {{N_{\rm{wg}}}} \right)} - {\bf{R}}_{2,2}^{\left( {{N_{\rm{wg}}}} \right)}{s_{{N_{\rm{t}}}}}} \right\|_2^2} \le {d^2}-C_1^{\left( {{N_{\rm{wg}}}} \right)},
\end{equation}
such that all constellation symbols satisfying this condition are considered as candidate demodulation symbols for $s_{N_{\rm wg}}$.

Next, we focus on the detection of $s_k,k=N_{\rm wg}-1,N_{\rm wg}-2,\cdots,1$, under the assumption that estimates of the symbols $\{{\hat s}_{k+1}, \cdots, {\hat s}_{N_{\rm wg}}\}$ are already available. Accordingly, we formulate \eqref{eq:P1_objective} as
\begin{equation}
	\begin{split}
		&{\left\| {\left[ {\begin{array}{*{20}{c}}
						{{\bf{z}}_1^{\left( k \right)}}\\
						{{\bf{z}}_2^{\left( k \right)}}\\
						{{\bf{z}}_3^{\left( k \right)}}
				\end{array}} \right] - \left[ {\begin{array}{*{20}{c}}
						{{\bf{R}}_{1,1}^{\left( k \right)}}&{{\bf{R}}_{1,2}^{\left( k \right)}}&{{\bf{R}}_{1,3}^{\left( k \right)}}\\
						{{\bf{0}}_{2,1}^{\left( k \right)}}&{{\bf{R}}_{2,2}^{\left( k \right)}}&{{\bf{R}}_{2,3}^{\left( k \right)}}\\
						{{\bf{0}}_{3,1}^{\left( k \right)}}&{{\bf{0}}_{3,2}^{\left( k \right)}}&{{\bf{R}}_{3,3}^{\left( k \right)}}
				\end{array}} \right]\left[ {\begin{array}{*{20}{c}}
						{{\bf{s}}_1^{\left( k \right)}}\\
						{{s_k}}\\
						{\hat{\bf{s}}_2^{\left( k \right)}}
				\end{array}} \right]} \right\|_2^2}\\
		=& \underbrace{{\left\| {{\bf{z}}_1^{\left( k \right)} - {\bf{R}}_{1,1}^{\left( k \right)}{\bf{s}}_1^{\left( k \right)} - {\bf{R}}_{1,2}^{\left( k \right)}{s_k} - {\bf{R}}_{1,3}^{\left( k \right)}{\hat{\bf s}}_2^{\left( k \right)}} \right\|_2^2}}_{T_1^{\left(k\right)}} \\
		&+ {\left\| {{\bf{z}}_2^{\left( k \right)} - {\bf{R}}_{2,2}^{\left( k \right)}{s_k} - {\bf{R}}_{2,3}^{\left( k \right)}{\hat{\bf s}}_2^{\left( k \right)}} \right\|_2^2} + \underbrace{{\left\| {{\bf{z}}_3^{\left( k \right)} - {\bf{R}}_{3,3}^{\left( k \right)}{\hat{\bf s}}_2^{\left( k \right)}} \right\|_2^2}}_{C_2^{\left(k\right)}},
	\end{split}
\end{equation}
where ${{\bf{z}}_1^{\left( k \right)}}\in {\mathbb C}^{k-1}$, ${{\bf{z}}_2^{\left( k \right)}}\in {\mathbb C}$, ${{\bf{z}}_3^{\left( k \right)}}\in {\mathbb C}^{N_{\rm r}-k}$, ${{\bf{R}}_{1,1}^{\left( k \right)}}\in{\mathbb C}^{\left(k-1\right)\times\left(k-1\right)}$, ${{\bf{R}}_{1,2}^{\left( k\right)}}\in{\mathbb C}^{\left(k-1\right)\times k}$, ${{\bf{R}}_{1,3}^{\left( k\right)}}\in{\mathbb C}^{\left(k-1\right)\times \left(N_{\rm wg}-k\right)}$, ${{\bf{0}}_{2,1}^{\left( k \right)}}\in {\mathbb C}^{1\times\left({k-1}\right)}$, ${{\bf{R}}_{2,2}^{\left( k \right)}}\in{\mathbb C}$, ${{\bf{R}}_{2,3}^{\left(k\right)}}\in{\mathbb C}^{1\times \left(N_{\rm wg}-k\right)}$, ${{\bf{0}}_{3,1}^{\left( k \right)}}\in {\mathbb C}^{\left(N_{\rm r}-k\right)\times\left({k-1}\right)}$, ${{\bf{0}}_{3,2}^{\left( k \right)}}\in {\mathbb C}^{\left({N_{\rm r}-k}\right)\times 1}$, ${{\bf{R}}_{3,3}^{\left( k\right)}}\in{\mathbb C}^{\left({N_{\rm r}-k}\right)\times \left(N_{\rm wg}-k\right)}$, ${{\bf{s}}_1^{\left( k \right)}}\in{\mathcal S}^{k-1}$ and ${{\bf{s}}_2^{\left( k \right)}}\in{\mathbb C}^{N_{\rm wg}-k}$.  The lower bound on ${T_1^{\left(k\right)}}$ can be obtained as
\begin{equation}
	C_1^{\left(k \right)} = {\left\| {{\bf{z}}_1^{\left( k \right)} - {\bf{R}}_{1,1}^{\left( k \right)}{\tilde{\bf s}}_1^{\left( k \right)} - {\bf{R}}_{1,2}^{\left( k \right)}{s_k} - {\bf{R}}_{1,3}^{\left( k \right)}{\hat{\bf s}}_2^{\left( k \right)}} \right\|_2^2},
\end{equation}
where ${\tilde{\bf s}}_1^{\left( k \right)}$ is the solution of
\begin{subequations}
\begin{align}
	&\left( {{\rm P}2\text{-}k} \right)\notag\\ &\mathop {\min }\limits_{{\bf{s}}_1^{\left( k \right)} \in {{\mathbb C}^{{k-1}}}} {\left\| {{\bf{z}}_1^{\left( k \right)} - {\bf{R}}_{1,1}^{\left( k \right)}{\bf{s}}_1^{\left( k \right)} - {\bf{R}}_{1,2}^{\left( k \right)}{s_k} - {\bf{R}}_{1,3}^{\left( k \right)}{\hat{\bf s}}_2^{\left( k \right)}} \right\|_2^2}\\
	&s.t.\quad \min\left(\Re\{\mathcal{S}\}\right)\cdot\mathbf{1}\leq \Re\left({{\bf{s}}_1^{\left( k \right)}}\right)\leq\max\left(\Re\{\mathcal{S}\}\right)\cdot\mathbf{1},\\
	&\phantom{s.t.\quad}\min\left(\Im\{\mathcal{S}\}\right)\cdot\mathbf{1}\leq \Im\left({{\bf{s}}_1^{\left( k \right)}}\right)\leq\max\left(\Im\{\mathcal{S}\}\right)\cdot\mathbf{1}.
\end{align}
\end{subequations}

Finally, the scope of the search can be further obtained as
\begin{equation}\label{eq:scope_s_k}
	{\left\| {{\bf{z}}_2^{\left( k \right)} - {\bf{R}}_{2,2}^{\left( k \right)}{s_k} - {\bf{R}}_{2,3}^{\left( k \right)}{\hat{\bf s}}_2^{\left( k \right)}} \right\|_2^2} \le {d^2}-C_1^{\left( k \right)}-C_2^{\left( k \right)}.
\end{equation}

Through an iterative procedure, we can determine the set of all candidates satisfying condition \eqref{eq:sd_constraint}, which is denoted by $\mathcal{C}_{\mathcal I}$. Accordingly, the optimum solution for a specific ${\mathcal I}$ can be obtained by
\begin{equation}
	{\hat{\bf{s}}_{\mathcal I}}=\arg \mathop {\min }\limits_{{{\bf{s}}} \in \mathcal{C}_{\mathcal I}} {\left\| {{\bf{z}}_{\mathcal I} - {{\bf{R}}_{\mathcal I}}{{\bf{s}}}} \right\|_2^2}.
\end{equation}

After the layered search is completed for a specific pattern $\mathcal{I}$, we obtain the conditional ML solution $\hat{\mathbf{s}}_{\mathcal{I}}$. This process is repeated for all possible activation patterns $\mathcal{I} \in \mathbb{I}$. Finally, the activation pattern $\hat{{\mathcal I}}$ is determined by minimizing the ML metric as
\begin{equation}
	\label{eq:final_solution}
	\hat{\mathcal{I}} = \arg \min_{\mathcal{I} \in \mathbb{I}} \left\| \mathbf{z}_{\mathcal{I}} - \mathbf{R}_{\mathcal{I}} \hat{\mathbf{s}}_{\mathcal{I}} \right\|_2^2,
\end{equation}
and the APM symbols are demodulated as ${\hat{\bf{s}}_{\hat{\mathcal I}}}$.
In a nutshell, the proposed BO-SD algorithm is summarized in Algorithm~\ref{alg:bo_sd}.
\begin{algorithm}[htbp]
	\renewcommand{\algorithmicrequire}{\textbf{Input:}}
	\renewcommand{\algorithmicensure}{\textbf{Output:}}
	\caption{Low-complexity BO-SD detector}
	\label{alg:bo_sd}
	\begin{algorithmic}[1]
		\REQUIRE $\mathbf{y}$, $\mathbf{H}$.
		\ENSURE $\hat{\mathcal{I}}$, ${\hat{\mathbf{s}}}_{\hat{\mathcal{I}}}$.
		\FOR{each index pattern ${\mathcal I}\in \mathbb I$}
		\STATE Construct $\mathbf{H}_{\mathcal{I},{\rm eq}}$ as \eqref{eq:sys_model_equivalent}, and compute its QR decomposition as $\mathbf{H}_{\mathcal{I},{\rm eq}} = \mathbf{Q}_{\mathcal{I}} \mathbf{R}_{\mathcal{I}}$;
		\STATE Compute transformed received vector as $\mathbf{z} = \mathbf{Q}_{\mathcal{I}} ^H \mathbf{y}$;
		\STATE Obtain $\tilde{\mathbf{s}}_{\mathcal{I}}$ by solving $\left( {{\rm P}1} \right)$;
		\STATE Set the radius as $d =  {\left\| {{\bf{z}} - {{\bf{H}}_{{\mathcal I},{\rm eq}}}\tilde{\mathbf{s}}_{\mathcal{I}}} \right\|_2}$;
		\FOR{$k=N_{\rm wg}$ to $1$}
		\IF{$k=N_{\rm wg}$}
		\STATE Obtain ${\bf{\tilde s}}_1^{\left( {{N_{\rm{wg}}}} \right)}$ by solving $\left( {{\rm P}2\text{-}N_{\rm wg}} \right)$;
		\STATE Obtain candidate symbols for ${\hat s}_{N_{\rm wg}}$ satisfying \eqref{eq:scope_s_nwg};
		\ELSIF{$k<N_{\rm wg}$}
		\STATE Obtain ${\bf{\tilde s}}_1^{k}$ by solving $\left( {{\rm P}2\text{-}k} \right)$;
		\STATE Obtain candidate symbols for ${\hat s}_k$ satisfying \eqref{eq:scope_s_k};
		\ENDIF
		\ENDFOR
		\ENDFOR
		\STATE ${\hat{\bf{s}}_{\mathcal I}}=\arg \mathop {\min }\limits_{{{\bf{s}}_{\mathcal I}} \in \mathcal{C}_{\mathcal I}} {\left\| {{\bf{z}} - {{\bf{R}}_{\mathcal I}}{{\bf{s}}_{\mathcal I}}} \right\|_2^2}$;
		\STATE 	$\hat{{\mathcal I}}=\arg \mathop {\min }\limits_{{{{\mathcal I}}} \in {\mathbb I}} {\left\| {{\bf{z}} - {{\bf{R}}_{\mathcal I}}{\hat{\bf{s}}_{\mathcal I}}} \right\|_2^2}$.
	\end{algorithmic}
\end{algorithm}

\subsection{Complexity Analysis}\label{sec:complexity}
The overall computational complexity of the proposed BO-SD detector is dominated by the number of legitimate PA position patterns, which is given by $\left(\left\lfloor \binom{N_{\rm t}}{N_{\rm a}} \right\rfloor_{2^p}\right)^{N_{\rm wg}}$, and the complexity of the detection process invoked for each pattern. 
According to \cite{complexity}, the algorithm associated with each PA position pattern imposes a computational complexity on the order of ${\mathcal O}\left(N_{\rm wg}^6\right)$. 
Consequently, the total computational complexity of the proposed BO-SD detector can be approximated as $C_{\mathrm{BO-SD}} = \mathcal{O}\left( \left(\left\lfloor \binom{N_{\rm t}}{N_{\rm a}} \right\rfloor_{2^p}\right)^{N_{\rm wg}}N_{\mathrm{wg}}^6 \right)$.

A salient advantage of the approach is that its complexity is independent of the constellation size~$M$. 
This stands in stark contrast to its optimal ML counterpart, whose complexity is given by $C_{\mathrm{ML}} = \mathcal{O}\left(\left(\left\lfloor \binom{N_{\rm t}}{N_{\rm a}} \right\rfloor_{2^p}\right)^{N_{\rm wg}}M^{N_{\mathrm{wg}}} \right)$, which escalates exponentially with the modulation order~$M$.
Therefore, the proposed BO-SD algorithm is capable of achieving a substantial complexity reduction, particularly for high-order modulation schemes, while attaining an optimal ML performance.

\section{Performance Analysis of the PA-IM System}\label{sec:performance_analysis}
Having established the PA-IM transmission framework and a practical method for signal detection, a rigorous theoretical understanding of its performance becomes essential. This section, therefore, is dedicated to the performance analysis of the PA-IM system, culminating in the derivation of an analytical upper bound on the BER, which will serve as both a performance benchmark and a foundation for further optimization.

An upper bound on the BER can be established by invoking the well-known union bound~\cite{upperbound}, which is expressed as
\begin{equation}\label{eq:union_bound}
	P_b \le \frac{1}{2^\eta} \sum_{i=1}^{2^\eta} \sum_{j=1}^{2^\eta} \frac{P(\mathbf{x}_i \to \mathbf{x}_j) n_{i,j}}{\eta},
\end{equation}
where $\eta$ is the spectral efficiency defined in Eq.~(\ref{eq:spectral_efficiency}). The term $P(\mathbf{x}_i \to \mathbf{x}_j)$ represents the pairwise error probability (PEP), signifying the probability of erroneously detecting the vector $\mathbf{x}_j$ given that $\mathbf{x}_i$ was transmitted. Finally, $n_{i,j}$ quantifies the number of bits in error between the information sequences corresponding to $\mathbf{x}_i$ and $\mathbf{x}_j$. A detailed derivation of the PEP is provided in the sequel.

The conditional PEP, given the channel realization $\mathbf{H}$, can be obtained as \cite{SM2}
\begin{align}\label{eq:conditional_pep}
	& P\left( \left. \mathbf{x}_i \to \mathbf{x}_j \right| \mathbf{H} \right) \nonumber \\
	& = P\left( \left\| \mathbf{y} - \sqrt{\rho} \mathbf{H} \mathbf{x}_j \right\|_2^2 \le \left\| \mathbf{y} - \sqrt{\rho} \mathbf{H} \mathbf{x}_i \right\|_2^2 \right) \nonumber \\
	& = P\left( \left\| \sqrt{\rho} \mathbf{H} (\mathbf{x}_i - \mathbf{x}_j) + \mathbf{n} \right\|_2^2 \le \left\| \mathbf{n} \right\|_2^2 \right) \nonumber \\
	& = Q\left( \sqrt{\frac{\rho \left\| \mathbf{H} \mathbf{\Delta}_{i,j} \right\|_2^2}{2N_0}} \right),
\end{align}
where the error vector is defined as $\mathbf{\Delta}_{i,j} \triangleq \mathbf{x}_i - \mathbf{x}_j$ and $Q(\cdot)$ is the Gaussian Q-function. 
\begin{remark}\label{re:positions}
As evident in \eqref{eq:conditional_pep}, the BER performance significantly depends on the receive SNR, which means that the candidate PA positions should be established in close proximity to the receiver to mitigate large-scale fading.
\end{remark}

By averaging the conditional PEP in Eq.~(\ref{eq:conditional_pep}) over the random channel statistics, the unconditional PEP can be obtained. More specifically, upon defining the random variable $\gamma_{i,j} \triangleq \| \mathbf{H} \mathbf{\Delta}_{i,j} \|_2^2$, the unconditional PEP is given by
\begin{equation}\label{eq:unconditional_pep}
	P(\mathbf{x}_i \to \mathbf{x}_j) = \int_0^{\infty} Q\left(\sqrt{\frac{\rho \gamma_{i,j}}{2N_0}}\right) p_{\gamma_{i,j}}(\gamma) d\gamma,
\end{equation}
where $p_{\gamma_{i,j}}(\gamma)$ is the probability density function (PDF) of $\gamma_{i,j}$. By invoking the alternative integral form of the Gaussian Q-function \cite{Qfunction}, $Q(x) = \frac{1}{\pi} \int_0^{\pi/2} \exp(-\frac{x^2}{2\sin^2\theta}) d\theta$, we arrive at
\begin{equation}\label{eq:unconditional_pep_mgf}
	P(\mathbf{x}_i \to \mathbf{x}_j) = \frac{1}{\pi} \int_0^{\pi/2} M_{\gamma_{i,j}}\left(-\frac{\rho}{4N_0\sin^2\theta}\right) d\theta,
\end{equation}
where $M_{\gamma_{i,j}}(s) \triangleq \int_0^{\infty} e^{s\gamma} p_{\gamma_{i,j}}(\gamma) d\gamma$ is the moment generating function (MGF) of $\gamma_{i,j}$. Moreover, the random variable $\gamma_{i,j}$ can be expressed in a Hermitian quadratic form as
\begin{equation}\label{eq:quadratic_form}
	\begin{split}
		\gamma_{i,j} &= \underbrace {{\rm{vec}}{{\left( {{{{\mathbf{H}}}^H}} \right)}^H}}_{{{\mathbf{u}}^H}}\underbrace {\left( {{{\mathbf{I}}_{{N_r}}} \otimes {\mathbf{\Delta }}_{i,j}{{\mathbf{\Delta }}^H_{i,j}}} \right)}_{{\mathbf Q}_{i,j}}\underbrace {{\rm{vec}}\left( {{{{\mathbf{H}}}^H}} \right)}_{\mathbf{u}} \\ 
		&= {{\mathbf{u}}^H}{\mathbf{Q}}_{i,j}{\mathbf{u}}. 
	\end{split}
\end{equation}

By applying the result for the characteristic function of Hermitian quadratic forms in complex normal variables presented in \cite{MGF}, the MGF of $\gamma_{i,j}$ can be expressed as
\begin{equation}\label{eq:mgf_of_gamma}
	 		{M_{\gamma_{i,j}} }\left( s \right) = \frac{{\exp \left\{ { s {{{\bf{\bar u}}}^H}{\bf{Q}}_{i,j} { {{\left( {{\bf{I}} - s{{\bf{C}}_{\bf u}}{\bf{Q}}_{i,j}} \right)}^{ - 1}}}{\bf{\bar u}}} \right\}}}{{\left| {{\bf{I}} - s{{\bf{C}}_{\bf u}}{\bf{Q}}_{i,j}} \right|}},
\end{equation}
where ${\mathbf{\bar u}} = {\mathbb E}\left\{ {\mathbf{u}} \right\}$ is the vector of the means and ${{\mathbf{C}}_{\mathbf{u}}} = {\mathbb E}\left\{ {\left( {{\mathbf{u}} - {\mathbf{\bar u}}} \right){{\left( {{\mathbf{u}} - {\mathbf{\bar u}}} \right)}^H}} \right\}$ is the covariance matrix. By plugging \eqref{eq:mgf_of_gamma} into \eqref{eq:unconditional_pep_mgf}, \eqref{eq:pep_nonclosed_form} shown at the top of the next page is obtained.

\begin{figure*}[htbp]
	\normalsize
	\begin{equation}\label{eq:pep_nonclosed_form}
		P\left( {{{\bf{x}}_i} \to {{\bf{x}}_j}} \right) = \frac{1}{\pi }\int\limits_0^{\pi /2} {\frac{{\exp \left\{ { -\frac{\rho }{{4{N_0}{{\sin }^2}\theta }}{{{\bf{\bar u}}}^H}{\bf{Q}}_{i,j}{{{\left( {{\bf{I}} + \frac{\rho }{{4{N_0}{{\sin }^2}\theta }}{{\bf{C}}_{\bf{u}}}{\bf{Q}}} \right)}^{ - 1}}}{\bf{\bar u}}} \right\}}}{{\left| {{\bf{I}} + \frac{\rho }{{4{N_0}{{\sin }^2}\theta }}{{\bf{C}}_{\bf{u}}}{\bf{Q}}} \right|}}d\theta } .
	\end{equation}
	\hrulefill
	\vspace*{4pt}
\end{figure*}

Now, we consider the derivation of ${\mathbf{\bar u}}$ and ${{\mathbf{C}}_{\mathbf{u}}}$.
Substituting \eqref{eq:subchannel_matrix} and \eqref{eq:overall_channel_matrix} into \eqref{eq:quadratic_form} yields \eqref{eq:u_vector}, shown at the top of the next page.
\begin{figure}[htbp]
	\normalsize
	\begin{small}
	\begin{equation}\label{eq:u_vector}
	\begin{split}
		{\bf{u}} = {{\bf{K}}_{{N_r}{N_t}}}\left[ {\begin{array}{*{20}{c}}
				{{\rm{vec}}\left( {{\bf{B}}_1^*} \right) \odot \left[ {{\rm{vec}}\left( {{\bf{K}}_{{\rm{LoS}},1}^*} \right) \odot {\rm{vec}}\left( {{\bf{\bar H}}_1^*} \right) + {\rm{vec}}\left( {{\bf{K}}_{{\rm{NLoS}},1}^*} \right) \odot {\rm{vec}}\left( {{\bf{\tilde H}}_1^*} \right)} \right] \odot {\rm{vec}}\left( {{\bf{\Gamma }}_1^*} \right)}\\
				{{\rm{vec}}\left( {{\bf{B}}_2^*} \right) \odot \left[ {{\rm{vec}}\left( {{\bf{K}}_{{\rm{LoS}},2}^*} \right) \odot {\rm{vec}}\left( {{\bf{\bar H}}_2^*} \right) + {\rm{vec}}\left( {{\bf{K}}_{{\rm{NLoS}},2}^*} \right) \odot {\rm{vec}}\left( {{\bf{\tilde H}}_2^*} \right)} \right] \odot {\rm{vec}}\left( {{\bf{\Gamma }}_2^*} \right)}\\
				\vdots \\
				{{\rm{vec}}\left( {{\bf{B}}_{{N_{{\rm{wg}}}}}^*} \right) \odot \left[ {{\rm{vec}}\left( {{\bf{K}}_{{\rm{LoS}},{N_{{\rm{wg}}}}}^*} \right) \odot {\rm{vec}}\left( {{\bf{\bar H}}_{{N_{{\rm{wg}}}}}^*} \right) + {\rm{vec}}\left( {{\bf{K}}_{{\rm{NLoS}},{N_{{\rm{wg}}}}}^*} \right) \odot {\rm{vec}}\left( {{\bf{\tilde H}}_{{N_{{\rm{wg}}}}}^*} \right)} \right] \odot {\rm{vec}}\left( {{\bf{\Gamma }}_{{N_{{\rm{wg}}}}}^*} \right)}
		\end{array}} \right].
	\end{split}
	\end{equation}
	\end{small}
	\hrulefill
	\vspace*{4pt}
\end{figure}
 In \eqref{eq:u_vector}, ${\bf{K}}_{{N_{\rm r}}{N_{\rm t}}}$ is the commutation matrix and defined as
\begin{equation}\label{eq:commutation_matrix}
	{{\bf{K}}_{{N_{\rm r}}{N_{\rm t}}}} \triangleq \sum\limits_{j = 1}^{{N_{\rm t}}} {\left( {{\bf{e}}_j^T \otimes {{\bf{I}}_{N_{\rm r}}} \otimes {{\bf{e}}_j}} \right)},
\end{equation}
where ${\bf e}_j$ is the $j$-th column vector of the $N_{\rm t}\times N_{\rm t}$ identity matrix.

Thus, ${\bf{\bar u}}$ can be formulated as \eqref{eq:mean_vector}
\begin{figure*}[htbp]
	\normalsize
	\begin{equation}\label{eq:mean_vector}
		\begin{aligned}
				{\bf{\bar u}} = \mathbb{E}\left\{ {{\rm{vec}}\left( {{{\bf{H}}^H}} \right)} \right\}
				= {{\bf{K}}_{{N_r}{N_t}}}\left[ {\begin{array}{*{20}{c}}
						{{\rm{vec}}\left( {{\bf{B}}_1^*} \right) \odot {\rm{vec}}\left( {{\bf{K}}_{{\rm{LoS}},1}^*} \right) \odot {\rm{vec}}\left( {{\bf{\tilde H}}_1^*} \right) \odot {\rm{vec}}\left( {{\bf{\Gamma }}_1^*} \right)}\\
						{{\rm{vec}}\left( {{\bf{B}}_2^*} \right) \odot {\rm{vec}}\left( {{\bf{K}}_{{\rm{LoS}},2}^*} \right) \odot {\rm{vec}}\left( {{\bf{\tilde H}}_2^*} \right) \odot {\rm{vec}}\left( {{\bf{\Gamma }}_2^*} \right)}\\
						\vdots \\
						{{\rm{vec}}\left( {{\bf{B}}_{{N_{{\rm{wg}}}}}^*} \right) \odot {\rm{vec}}\left( {{\bf{K}}_{{\rm{LoS}},{N_{{\rm{wg}}}}}^*} \right) \odot {\rm{vec}}\left( {{\bf{\tilde H}}_{{N_{{\rm{wg}}}}}^*} \right) \odot {\rm{vec}}\left( {{\bf{\Gamma }}_{{N_{{\rm{wg}}}}}^*} \right)}
				\end{array}} \right],
		\end{aligned}
	\end{equation}
	\hrulefill
	\vspace*{4pt}
\end{figure*}
and ${{\mathbf{C}}_{\mathbf{u}}}$ is given by
\begin{equation}\label{eq:convariance_matrix}
		\begin{aligned}
			{{\bf{C}}_{\bf{u}}} &= \mathbb{E}\left\{ {{\rm{vec}}\left( {{{{\bf{H}}}^H}} \right){\rm{vec}}{{\left( {{{{\bf{ H}}}^H}} \right)}^H}} \right\} - {\bf{\bar u}}{{{\bf{\bar u}}}^H}\\
			&= {{\bf{K}}_{{N_{\rm r}}{N_{\rm t}}}}\cdot\mathrm{blkdiag}(\mathbf{C}_{\mathbf{u},1}, \dots, \mathbf{C}_{\mathbf{u},N_{\rm wg}})\cdot{{\bf{K}}_{{N_{\rm r}}{N_{\rm t}}}^T},
		\end{aligned}
\end{equation}
where $\mathbf{C}_{\mathbf{u},n}$ is given in \eqref{eq:coviance_block_matrix} in the next page.
\begin{figure*}[htbp]
	\normalsize
\begin{equation}\label{eq:coviance_block_matrix}
	\mathbf{C}_{\mathbf{u},n} ={{\rm{vec}}\left( {{\bf{B}}_{n}^*} \right){\rm{vec}}{{\left( {{\bf{B}}_{n}^*} \right)}^H} \odot {\rm{vec}}\left( {{\bf{K}}_{{\rm{LoS}},n}^*} \right){\rm{vec}}{{\left( {{\bf{K}}_{{\rm{LoS}},n}^*} \right)}^H} \odot {\bf{I}} \odot {\rm{vec}}\left( {{\bf{\Gamma }}_n^*} \right){\rm{vec}}{{\left( {{\bf{\Gamma }}_n^*} \right)}^H}}.
\end{equation}
	\hrulefill
\vspace*{4pt}
\end{figure*}

Although \eqref{eq:pep_nonclosed_form} does not have a closed-form solution, a tight approximation can be derived with the aid of the approximation of the Q-function, which can be expressed as 
\begin{equation}\label{eq:q_function}
	Q\left( x \right) \approx \frac{1}{{12}}{e^{ - \frac{{{x^2}}}{2}}} + \frac{1}{4}{e^{ - \frac{{2{x^2}}}{3}}},x \ge 0.
\end{equation}
Then, the value of $P\left( {{{\mathbf{x}}_i} \to {{\mathbf{x}}_j}} \right)$ can be approximated as
\begin{equation}\label{eq:pep_closed_form}
	\begin{split}
			&P\left( {{{\bf{x}}_i} \to {{\bf{x}}_j}} \right) \\
			\approx& \int\limits_0^\infty  {\left( {\frac{1}{{12}}{e^{ - \frac{{\rho \gamma }}{{4{N_0}}}}} + \frac{1}{4}{e^{ - \frac{{\rho \gamma }}{{3{N_0}}}}}} \right)} {f_{{\gamma _{i,j}}}}\left( \gamma  \right)d\gamma \\
			=& \frac{1}{{12}}{M_{{\gamma _{i,j}}}}\left( { - \frac{\rho }{{4{N_0}}}} \right) + \frac{1}{4}{M_{{\gamma _{i,j}}}}\left( { - \frac{\rho }{{3{N_0}}}} \right)\\
			=& \frac{1}{{12}}\frac{{\exp \left( { - \frac{\rho }{{4{N_0}}}{{{\bf{\bar u}}}^H}{{\bf{Q}}_{i,j}}{{\left( {{\bf{I}} + \frac{\rho }{{4{N_0}}}{{\bf{C}}_{\bf{u}}}{{\bf{Q}}_{i,j}}} \right)}^{ - 1}}{\bf{\bar u}}} \right)}}{{\left| {{\bf{I}} + \frac{\rho }{{4{N_0}}}{{\bf{C}}_{\bf{u}}}{{\bf{Q}}_{i,j}}} \right|}}\\
			&+ \frac{1}{4}\frac{{\exp \left( { - \frac{\rho }{{3{N_0}}}{{{\bf{\bar u}}}^H}{{\bf{Q}}_{i,j}}{{\left( {{\bf{I}} + \frac{\rho }{{3{N_0}}}{{\bf{C}}_{\bf{u}}}{{\bf{Q}}_{i,j}}} \right)}^{ - 1}}{\bf{\bar u}}} \right)}}{{\left| {{\bf{I}} + \frac{\rho }{{3{N_0}}}{{\bf{C}}_{\bf{u}}}{{\bf{Q}}_{i,j}}} \right|}}.
	\end{split}
\end{equation}

\section{Transmit Precoding Design for BER Performance Enhancement}\label{sec:precoding_design}
The performance analysis in the preceding section provides crucial insights into the factors governing the system's BER. To proactively enhance system reliability from the transmitter's side, this section leverages these insights to design a novel transmit precoding scheme aimed at minimizing the derived BER bound.
The proposed PA-IM system is similar to conventional MIMO systems employing hybrid beamforming.
To elaborate, within a single waveguide, the signals radiated from the activated PAs are merely phase-shifted versions of each other. In other words, analog beamforming can be realized by acitvating pinching antennas at different locations, but precludes sophisticated intra-waveguide precoding or multi-stream transmission.
However, in a multi-waveguide configuration where each waveguide is associated with a dedicated RF chain, independent digital signal processing can be performed across the waveguides, which is similar to digital beamforming.
Therefore, transmit precoding may be readily applied across these waveguides to improve the system's BER performance provided that CSI is available at the transmitter \cite{channel_estimation1,channel_estimation2}.

\subsection{Problem Formulation}
Denote the linear diagonal transmit precoding matrix by $\mathbf{W} \in \mathbb{C}^{N_{\rm t} N_{\rm wg} \times N_{\rm t} N_{\rm wg}}$. The received signal vector can then be obtained as
\begin{equation}\label{eq:precoded_received_signal}
	\mathbf{y} = \sqrt{\rho} \mathbf{H} \mathbf{W} \mathbf{x} + \mathbf{n},
\end{equation}
where the precoding matrix $\mathbf{W}$ is structured to apply a distinct complex-valued weight $w_k$ to all $N_t$ candidate positions on the $k$-th waveguide, i.e.,
\begin{equation}
	\label{eq:precoder_structure}
	\mathbf{W} = \mathrm{diag}\left( w_1 \mathbf{1}_{N_{\rm t}}^T, w_2 \mathbf{1}_{N_{\rm t}}^T, \dots, w_{N_{\rm wg}} \mathbf{1}_{N_{\rm t}}^T \right).
\end{equation}
For the sake of preserving the total transmit power, a normalization constraint is imposed on the precoding weights, which can be expressed as
\begin{equation}
	\label{eq:power_constraint}
	\| \mathbf{W} \|_F^2 = N_{\rm t} \sum_{k=1}^{N_{\rm wg}} |w_k|^2 = N_{\rm t} N_{\rm wg}.
\end{equation}

To establish a design criterion for the precoder, we commence by deriving the conditional BER bound. By substituting the precoded signal model into the union bound in Eq.~(\ref{eq:union_bound}), we have
\begin{equation}
	P_{b|\mathbf{H}}(\mathbf{W}) \le \frac{1}{\eta 2^\eta} \sum_{i=1}^{2^\eta} \sum_{j=1}^{2^\eta} n_{i,j} Q\left( \sqrt{\frac{\rho \| \mathbf{H} \mathbf{W} \mathbf{\Delta}_{i,j} \|_2^2}{2N_0}} \right).
\end{equation}
Furthermore, by using the accurate approximation of the Q-function in \eqref{eq:q_function}, a tractable upper bound on the conditional BER, denoted by $P_{b|\mathbf{H}}^{\rm ub}\left(\mathbf{W}\right) $, can be formulated as Eq.~(\ref{eq:ber_upper_bound}) at the top of the next page.

\begin{figure*}[!t]
	\normalsize
	\begin{equation}\label{eq:ber_upper_bound}
		P_{b|\mathbf{H}}^{\rm ub}\left(\mathbf{W}\right) \triangleq \frac{1}{\eta 2^\eta} \sum_{i=1}^{2^\eta} \sum_{j=1}^{2^\eta} n_{i,j} \left[ \frac{1}{12} \exp\left(-\frac{\rho \| \mathbf{H} \mathbf{W} \mathbf{\Delta}_{i,j} \|_2^2}{4N_0}\right) + \frac{1}{4} \exp\left(-\frac{\rho \| \mathbf{H} \mathbf{W} \mathbf{\Delta}_{i,j} \|_2^2}{3N_0}\right) \right].
	\end{equation}
	\hrulefill
	\vspace*{4pt}
\end{figure*}

The design of the transmit precoder that minimizes this BER upper bound can then be cast as the following optimization problem:
\begin{subequations}
	\label{eq:optimization_problem_P3}
	\begin{align}
		\left( {{\rm P}3} \right)\quad&\min_{\substack{\bf W}}\quad {P_{b\left| {\bf{H}} \right.}^{\rm ub}\left({\bf W}\right) }\\
		&s.t.\quad \left\| {\bf{W}} \right\|_F^2  = {{N_{\rm t}}{N_{\rm wg}}}.
	\end{align}
\end{subequations}

\subsection{Precoder Design via Manifold Optimization}
The optimization problem $\left( {{\rm P}3} \right)$ is non-convex and challenging to solve directly. However, by defining the vector of precoding weights as $\mathbf{w} = [w_1, w_2, \dots, w_{N_{\rm wg}}]^T \in \mathbb{C}^{N_{\rm wg}}$, the optimization can be reformulated. Specifically, we have $\| \mathbf{H} \mathbf{W} \mathbf{\Delta}_{i,j} \|_2^2 = \| \mathbf{H} \mathbf{D}_{i,j} \mathbf{w} \|_2^2$, where $\mathbf{D}_{i,j} \triangleq \mathrm{diag}(\mathbf{d}_{i,j,1}, \dots, \mathbf{d}_{i,j,N_{\rm wg}})$ and ${{\bf{\Delta }}_{i,j}} = {\left[ {\begin{array}{*{20}{c}}
			{{\bf{d}}_{i,j,1}^T}&{{\bf{d}}_{i,j,2}^T}& \cdots &{{\bf{d}}_{i,j,{N_{{\rm{wg}}}}}^T}
	\end{array}} \right]^T}$. The problem can be recast equivalently as follows:
\begin{subequations}
	\label{eq:optimization_problem_w}
	\begin{align}
		\left( {{\rm P}3'} \right)\quad & \min_{\mathbf{w}} \quad f(\mathbf{w}) \\
		&s.t. \quad \| \mathbf{w} \|_2^2 = N_{\rm wg},
	\end{align}
\end{subequations}
where
\begin{equation}\label{eq:definition_of_f}
	\begin{split}
		f\left({\bf w}\right) =& \sum\limits_{i = 1}^{{2^\eta}} {\sum\limits_{j = 1}^{{2^\eta}} \frac{{n_{i,j}}}{{\eta{2^\eta}}}{\left[ {\frac{1}{{12}}\exp \left( { - \frac{{\rho\left\| {{\bf{H}}{\bf D}_{i,j}{{\bf{w}}}} \right\|_2^2}}{{4{N_0}}}} \right)} \right.} } \\
		&+\left. { \frac{1}{4}\exp \left( { - \frac{{\rho\left\| {{\bf{H}}{\bf D}_{i,j}{{\bf{w}}}} \right\|_2^2}}{{3{N_0}}}} \right)} \right].
	\end{split}
\end{equation}
Indeed, the feasible set for $\mathbf{w}$ is the complex sphere $\mathcal{M}^{N_{\rm wg}-1} = \{ \mathbf{w} \in \mathbb{C}^{N_{\rm wg}} : \mathbf{w}^H \mathbf{w} = N_{\rm wg} \}$, which can be viewed as a Riemannian submanifold of $\mathbb{C}^n$. This geometric structure allows us to invoke manifold optimization to find a high-quality solution.

Specifically, a gradient descent algorithm can be developed on the Riemannian manifold. The Riemannian gradient of the objective function $f(\mathbf{w})$ at a point $\mathbf{w}$ is the orthogonal projection of its Euclidean gradient $\nabla f(\mathbf{w})$ onto the tangent space $T_{\mathbf{w}}\mathcal{M}^{N_{\rm wg}-1} = \{ \mathbf{v} \in \mathbb{C}^{N_{\rm wg}} : \Re\{\mathbf{v}^H \mathbf{w}\} = 0 \}$. More specifically, this projection is given by
\begin{equation}\label{eq:gradient_of_f}
	\mathrm{grad} f(\mathbf{w}) = \nabla f(\mathbf{w}) - \frac{\Re\{(\nabla f(\mathbf{w}))^H \mathbf{w}\}}{\| \mathbf{w} \|_2^2} \mathbf{w},
\end{equation}
where the Euclidean gradient $\nabla f(\mathbf{w})$ can be derived from the objective function, as detailed in Eq.~(\ref{eq:euclidean_gradient}) at the top of the next page.

\begin{figure*}[htbp]
	\normalsize
	\begin{small}
	\begin{equation}\label{eq:euclidean_gradient}
		\nabla f\left( {\bf{w}} \right) = \sum\limits_{i = 1}^{{2^\eta }} {\sum\limits_{j = 1}^{{2^\eta }} {\frac{{{n_{i,j}}}}{{\eta {2^\eta }}}\left[ {\frac{{\rho {\bf{D}}_{i,j}^H{{\bf{H}}^H}{\bf{H}}{{\bf{D}}_{i,j}}{\bf{w}}}}{{24{N_0}}}\exp \left( { - \frac{{\rho \left\| {{\bf{H}}{{\bf{D}}_{i,j}}{\bf{w}}} \right\|_2^2}}{{4{N_0}}}} \right) + \frac{{\rho {\bf{D}}_{i,j}^H{{\bf{H}}^H}{\bf{H}}{{\bf{D}}_{i,j}}{\bf{w}}}}{{6{N_0}}}\exp \left( { - \frac{{\rho \left\| {{\bf{H}}{{\bf{D}}_{i,j}}{\bf{w}}} \right\|_2^2}}{{3{N_0}}}} \right)} \right]} }.
	\end{equation}
	\end{small}
	\hrulefill
\end{figure*}

Once the search direction, given by the negative Riemannian gradient, is determined, the next iteration point is found by a retraction operation. The retraction maps a tangent vector back onto the manifold. For the sphere, the retraction of a tangent vector $-\beta\operatorname{grad}f\left({\bf w}\right)$ at point ${\bf w} \in {\mathcal M}^{{N_{\rm t}}{N_{\rm wg}}-1} $ can be obtained as
\begin{equation}\label{eq:definition_of_retr}
	\mathrm{Retr}_{\mathbf{w}}(-\beta \cdot \mathrm{grad}f(\mathbf{w})) = \sqrt{N_{\rm wg}} \frac{\mathbf{w} - \beta \cdot \mathrm{grad}f(\mathbf{w})}{\| \mathbf{w} - \beta \cdot \mathrm{grad}f(\mathbf{w}) \|_2},
\end{equation}
where $\beta$ is the step size, which can be determined using a line search procedure such as the Armijo rule~\cite{Armijo}. Equipped with these components, a Riemannian gradient descent algorithm can be formulated to solve $\left( {{\rm P}3'} \right)$, as summarized in Algorithm~\ref{alg:manifold_opt}. By virtue of Theorem 4.3.1 of~\cite{convergence}, this algorithm is guaranteed to converge to a stationary point of the optimization problem. It is worth noting that since the objective function in (50) is non-convex, the algorithm converges to a stationary local minimum. However, as the simulation results will demonstrate, this approach effectively finds a high-quality precoding solution that yields substantial BER improvements.

\begin{algorithm}[htbp]
	\renewcommand{\algorithmicrequire}{\textbf{Input:}}
	\renewcommand{\algorithmicensure}{\textbf{Output:}}
	\caption{Gradient descent algorithm for transmit precoding matrix based on manifold optimization}
	\label{alg:manifold_opt}
	\begin{algorithmic}[1]
		\REQUIRE $\mathbb{X}$, $\mathbf{H}$, $\xi$.
		\ENSURE $\bf W$.
		\STATE{Obtain $\operatorname{grad}f\left({\bf w}_{0}\right)$ via \eqref{eq:gradient_of_f} and \eqref{eq:euclidean_gradient};}
		\STATE{$k=0$;}
		\REPEAT
		\STATE{Determine Armijo backtracking line search step size $\beta_k$;}
		\STATE{Determine the next point using retraction as ${\bf w}_{k+1}={\rm Retr}_{{\bf w}_k}\left(-\beta_k\operatorname{grad}f\left({\bf w}_k\right)\right)$;}
		\STATE{Determine Riemannian gradient $\operatorname{grad}f\left({\bf w}_{k+1}\right)$ according to \eqref{eq:gradient_of_f} and \eqref{eq:euclidean_gradient};}
		\STATE{$k\leftarrow k+1$}
		\UNTIL{Decrease of ${f\left({\bf w}\right) }$ is below $\xi$;}
		\STATE ${\bf W}={\rm diag}\left({\bf w}\right)\otimes {\bf I}_{N_{\rm t}}$.
	\end{algorithmic}
\end{algorithm}

\section{Simulation Results}\label{sec:simulation}

In this section, we present the performance results characterizing the proposed PA-IM system and validate the derived analytical upper bound on the BER. The results also illustrate the advantages of the proposed transmit precoding design.

\subsection{Simulation Setups and System Parameters}
For illustration purposes, we assume that the side length of the area is $D=500$ m, the noise power is $N_0=-90$ dBm, the carrier frequency is $f_c=3$ GHz and the effective refractive index of the waveguide $\eta_{\rm eff}=1.4$. The position of the receiver is set at $\left(400{\rm m},50{\rm m},1.5{\rm m}\right)$ in Fig. 1 and the corresponding ULA is aligned with the $x$-axis, with an inter-element spacing of half a wavelength to avoid mutual coupling effects. The setups for the proposed PA-IM system and conventional SM system are as follows.

\subsubsection{Conventional SM System}
The transmitter is located at the center of the square area in Fig. 1 with a height of $12.5$ m, whose antennas are aligned with the $x$-axis, with an inter-element spacing of half a wavelength.

\subsubsection{PA-IM System}
The transmitter is located at the origin of the square area in Fig. 1 with a height of $12.5$ m. $N_{\rm wg}$ waveguides are placed parallel to the $x$-axis and are equidistantly distributed along the $y$-axis. Candidate positions for the PAs on each waveguide are established around the leakage point closest to the user with a half-wavelength spacing to mitigate large-scale fading and enhance the received SNR, as analyzed in Remark \ref{re:positions} of Section \ref{sec:performance_analysis}. All configurations considered in this section are summarized in Table \ref{tab:my_data_professional}.

\begin{table}[htbp]
	\centering
	\renewcommand{\arraystretch}{1.5}
	\caption{Simulation Parameter Configuration for PA-IM Systems.} 
	\label{tab:my_data_professional}
	\begin{tabularx}{0.48\textwidth}{
			>{\bfseries}l
			*{6}{>{\centering\arraybackslash}X} 
		}
		\toprule 
		& \makecell{$\boldsymbol{N_{\rm t}}$} 
		& \makecell{$\boldsymbol{N_{\rm wg}}$} 
		& \makecell{$\boldsymbol{N_{\rm a}}$} 
		& \makecell{$\boldsymbol{N_{\rm r}}$} 
		& \makecell{$\boldsymbol{M}$}
		& \makecell{$\boldsymbol{P_{\rm t}}$ \\ \text{(dBm)}} \\ 
		\midrule 
		Fig. 2 & 4   & 1   & 1   & 2   & $-$ & $-$ \\ 
		Fig. 3(a) & 4   & 2   & 1   & 4   & $-$ & $-$ \\ 
		Fig. 3(b) & 4   & 1   & 1   & $-$ & $-$ & $-$ \\ 
		Fig. 4 & 8   & 1   & 1   & $-$   & $-$ & $-$ \\ 
		Fig. 5 & 4   & 1   & 2   & 1   & $-$ & $-$ \\ 
		Fig. 6 & 8   & $-$ & 1 & 2   & 2   & 20  \\ 
		\bottomrule
	\end{tabularx}
\end{table}

As for the parameters for the channel, the LoS component is set as $1$. Based on the 3GPP channel model in \cite{3gpp} and recommendations in \cite{LoS1,LoS2}, the possibility of having an LoS component between the $j$-th activated PA on the $n$-th waveguide and the $i$-th antenna of the receiver mainly depends on the distance $d_{i,j}^n$ between them. More specifically, the possibility of having an LoS component is
\begin{equation}
	P(\text{LoS}) = \begin{cases} 1 - \frac{d_{i,j}^n}{300}, & 0 < d_{i,j}^n < 300 \text{ m}, \\ 0, & d_{i,j}^n \ge 300 \text{ m.} \end{cases}
\end{equation}
Accordingly, the Rician factor can be calculated as \cite{LoS1,LoS2}
\begin{equation}
	K_{i,j}^n = \begin{cases} 10^{1.3 - 0.003d_{i,j}^n}, & \text{if LoS component exists}, \\ 0, & \text{if LoS component does not exist.} \end{cases}
\end{equation}

For the pathloss model, we apply the COST 321 WalfischIkegami model and the corresponding large-scale channel fading coefficient (in dB) between is given as \cite{LoS1,LoS2}
\begin{equation}
	\beta _{i,j}^n= \begin{cases} -30.18 - 26\log_{10}(d_{i,j}^n) + F_{i,j}^n, & K_{i,j}^n \neq 0, \\ -34.53 - 38\log_{10}(d_{i,j}^n) + F_{i,j}^n, & K_{i,j}^n = 0, \end{cases}
\end{equation}
where $F_{i,j}^n=\sqrt{\delta}a_{j}^n+\sqrt{1-\delta}b_{i}$ represents the correlated shadow fading coefficient, where $\delta=0.5$, $\sigma_{\rm sf}=8$, $a_j^n\sim {\mathcal N}\left(0,\sigma_{\rm sf}^2\right)$ and $b_i\sim {\mathcal N}\left(0,\sigma_{\rm sf}^2\right)$. The covariance functions of $a_j^n$ and $b_i$ are given by
\begin{equation}
	\mathbb{E} \{a_j^n a_{j'}^{n'}\} = 2^{-\frac{d_{\rm PA}(n,j;n',j')}{d_{\rm decorr}}} , \quad \mathbb{E} \{b_k b_{k'}\} = 2^{-\frac{d_{\rm r}(k,k')}{d_{\rm decorr}}} ,
\end{equation}
where $d_{\rm PA}(n,j;n',j')$ is the geographical distance between the $j$-th activated PA on the $n$-th waveguide and the $j'$-th activated PA on the $n'$-th waveguide, and $d_{\rm r}(k,k')$ is the geographical distance  between the $k$-th and $k'$-th antenna of the receiver, and $d_{\rm decorr}$ is a decorrelation  distance that depends on the environment, typically set as $d_{\rm decorr}=100$ m.

\subsection{Results and Discussions}

\begin{figure}[htbp]
	\centering
	\includegraphics[width=3.35in]{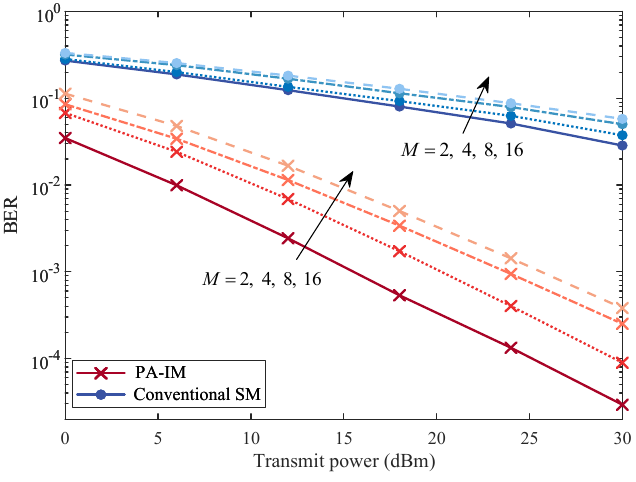}
	\caption{BER performances for the PA-IM and conventional SM systems.}
	\label{PAvsSM}
\end{figure}
Firstly, the BER performance of the proposed PA-IM system is compared to that of the conventional SM system in Fig.~\ref{PAvsSM}. For a fair comparison, the PA-IM system is equipped with $\left(N_{\rm t},N_{\rm wg},N_{\rm a},N_{\rm r}\right)=\left(4,1,1,2\right)$, while the conventional SM system employs four transmit antennas, one of which is activated during transmission, and two receive antennas. As expected, the PA-IM scheme realizes a substantial BER performance gain over its conventional SM counterpart. More specifically, for a spectral efficiency of 4 bits/s/Hz (i.e., $M=4$), the PA-IM system provides an SNR gain of approximately 30~dB at the BER of $10^{-1}$ compared to SM. This substantial gain is attributed to the PA-IM's capability of dynamically configuring its antenna positions. By proactively establishing a dominant LoS path, the system effectively mitigates large-scale fading and engineers a more favorable, quasi-deterministic channel, thereby ensuring a significantly higher received SNR than its counterpart.

\begin{figure}[!t]
	\centering
	\subfigure[]{\includegraphics[width=3.35in]{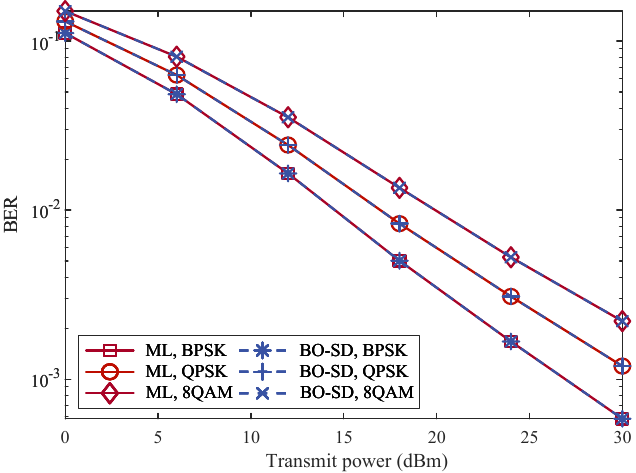}
		\label{fig_first_case}}
	\hfil
	\subfigure[]{\includegraphics[width=3.35in]{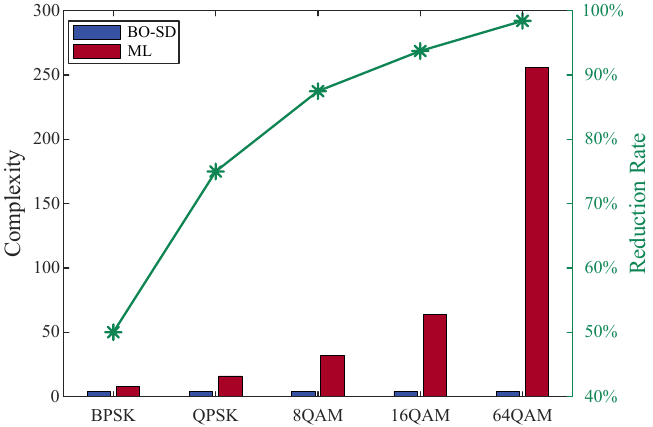}
		\label{fig_second_case}}
	\caption{ Performance results of PA-IM systems with different detectors. (a) BER. (b) Complexity.}
	\label{fig_sim}
\end{figure}

The performance results of the proposed BO-SD detector is evaluated in Fig.~\ref{fig_sim}, which are benchmarked against the optimal ML detector. It can be readily observed in Fig.~\ref{fig_first_case} that the performance of the BO-SD detector is identical with that of the ML detector across all modulation schemes. This confirms that the proposed BO-SD algorithm is capable of achieving optimal ML performance.
However, their computational complexities differ dramatically, which is a crucial aspect for practical implementation.
To quantify this, Fig.~\ref{fig_second_case} provides a comparison of the computational complexity for both detectors.
It is evident that the complexity of the ML detector escalates exponentially with the modulation order, rendering it prohibitive for high-order modulation schemes.
In stark contrast, the complexity of the BO-SD detector remains low and independent of the modulation order $M$, consistent with our analysis in Section~\ref{sec:complexity}.
Specifically, in the case of 64-QAM, the BO-SD algorithm achieves a complexity reduction of approximately $98.5\%$ compared to its ML counterpart.
This beneficial trade-off, i.e., attaining optimal BER performance at a fraction of the computational cost, underscores the practical viability and superiority of the proposed BO-SD detector, particularly for spectrally efficient systems employing higher-order modulation.

\begin{figure}[htbp]
	\centering
	\includegraphics[width=3.35in]{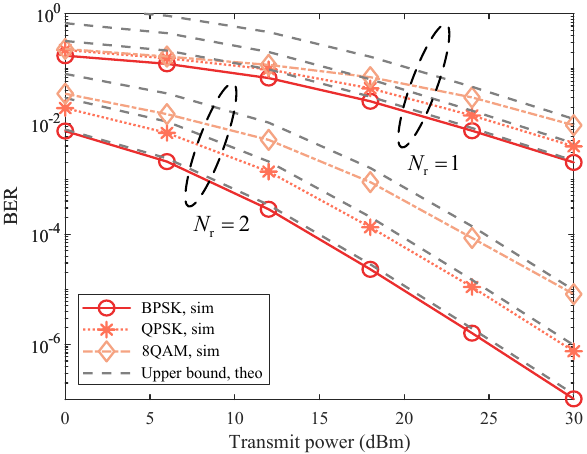}
	\caption{BER performances of PA-IM systems with different modulation orders.}
	\label{upperbound}
\end{figure}
Fig.~\ref{upperbound} investigates the BER performance of the proposed PA-IM system under different modulation orders, while the number of receiver antennas is set as $N_{\rm r} = 1$ and $N_{\rm r} = 2$. It can be readily observed that the simulation results in all configurations are in excellent agreement with the derived theoretical upper bounds, especially in the high-SNR region, which validates the accuracy of the theoretical analysis. Additionally, a significant performance enhancement is achieved by increasing the number of receiver antennas, which implies that the increased receive diversity effectively mitigates the effects of fading.

\begin{figure}[htbp]
	\centering
	\includegraphics[width=3.35in]{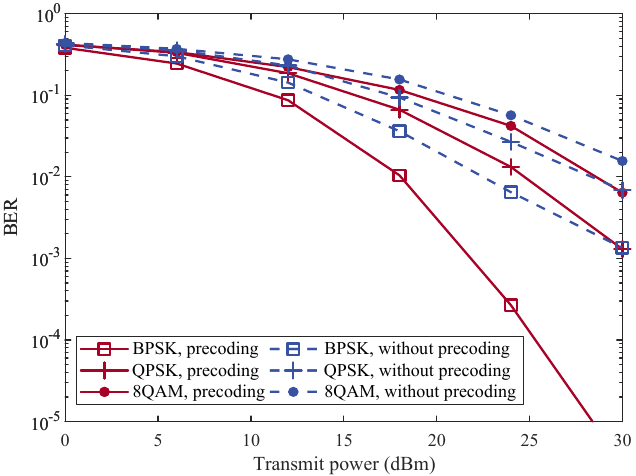}
	\caption{BER performances for the PA-IM systems with and without transmit precoding.}
	\label{precoding}
\end{figure}
The effectiveness of the proposed minimum-BER based precoding design is validated in Fig.~\ref{precoding}, which showcases its superior performance in the PA-IM system. For comparison, the performance of the PA-IM system without transmit precoding is included as a baseline. It is evident that the transmit precoding-aided scheme significantly outperforms its counterpart without transmit precoding across all considered modulation orders. For instance, the transmit precoding design achieves a remarkable SNR gain of nearly $9$~dB at the BER of approximately $10^{-3}$ for the BPSK case. This stems from the fact that the proposed transmit precoding algorithm, based on manifold optimization, aims at maximizing the Euclidean distance between received signals, thereby enhancing the system's resilience to the fading as well as noise.

\begin{figure}[htbp]
	\centering
	\includegraphics[width=3.35in]{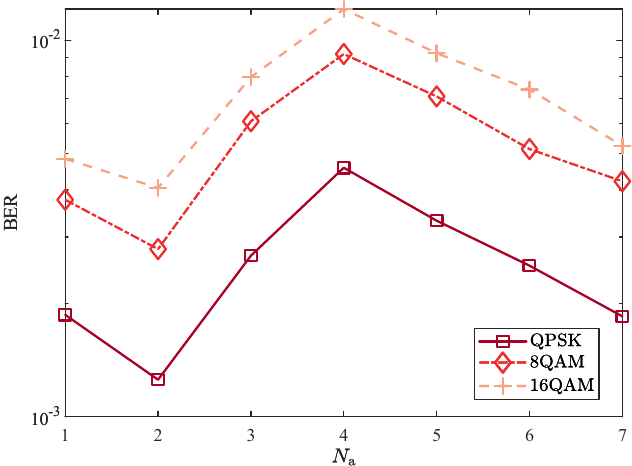}
	\caption{BER performances of PA-IM systems with different $N_{\rm a}$.}
	\label{Na}
\end{figure}
Fig.~\ref{Na} illustrates the impact of the number of activated PAs per waveguide, $N_{\rm a}$, on the BER performance. It can be observed that the BER performance does not improve monotonically with $N_{\rm a}$; instead, an optimal $N_{\rm a}$ exists for each scenario. This trend reveals a fundamental trade-off: while a higher $N_{\rm a}$ enhances spatial diversity against the fading, it also reduces the power allocated to each activated antenna under a fixed total transmit power per waveguide.

\section{Conclusion}\label{sec:conclusion}
In this paper, we proposed the novel PA-IM scheme, which effectively incorporates the principle of IM to enhance the spectral efficiency of PA systems without increasing hardware complexity. To tackle the prohibitive complexity of optimal ML detection at the receiver, a low-complexity BO-SD algorithm was conceived, which adaptively prunes the search space while preserving optimal ML performance. Furthermore, to further enhance the PA-IM system performance from the transmitter's perspective, a new transmit precoding method using manifold optimization was designed to minimize the BER by maximizing the minimum Euclidean distance between all received signals. Our simulation results demonstrated that the proposed PA-IM scheme, supported by our transceiver design, attains a significant performance gain over its conventional counterparts, and the overall BER of the system could be further improved by the proposed transmit precoding design.

\ifCLASSOPTIONcaptionsoff
  \newpage
\fi

\bibliographystyle{IEEEtran}      
\bibliography{IEEEabrv,main}

\begin{thebibliography}{10}
\providecommand{\url}[1]{#1}
\csname url@samestyle\endcsname
\providecommand{\newblock}{\relax}
\providecommand{\bibinfo}[2]{#2}
\providecommand{\BIBentrySTDinterwordspacing}{\spaceskip=0pt\relax}
\providecommand{\BIBentryALTinterwordstretchfactor}{4}
\providecommand{\BIBentryALTinterwordspacing}{\spaceskip=\fontdimen2\font plus
\BIBentryALTinterwordstretchfactor\fontdimen3\font minus
  \fontdimen4\font\relax}
\providecommand{\BIBforeignlanguage}[2]{{%
\expandafter\ifx\csname l@#1\endcsname\relax
\typeout{** WARNING: IEEEtran.bst: No hyphenation pattern has been}%
\typeout{** loaded for the language `#1'. Using the pattern for}%
\typeout{** the default language instead.}%
\else
\language=\csname l@#1\endcsname
\fi
#2}}
\providecommand{\BIBdecl}{\relax}
\BIBdecl

\bibitem{Shannon}
C.~E. Shannon, ``A mathematical theory of communication,'' \emph{Bell Syst.
  Tech. J.}, vol.~27, no.~3, pp. 379--423, Jul. 1948.

\bibitem{mMIMO}
E.~G. Larsson, O.~Edfors, F.~Tufvesson, and T.~L. Marzetta, ``Massive {MIMO}
  for next generation wireless systems,'' \emph{IEEE Commun. Mag.}, vol.~52,
  no.~2, pp. 186--195, Feb. 2014.

\bibitem{6G1}
Y.~Xiao, Z.~Ye, M.~Wu, H.~Li, M.~Xiao, M.-S. Alouini, A.~Al-Hourani, and
  S.~Cioni, ``Space-air-ground integrated wireless networks for {6G}: Basics,
  key technologies, and future trends,'' \emph{IEEE J. Sel. Areas Commun},
  vol.~42, no.~12, pp. 3327--3354, Dec. 2024.

\bibitem{6G2}
P.~Yang, Y.~Xiao, M.~Xiao, and S.~Li, ``6{G} wireless communications: Vision
  and potential techniques,'' \emph{IEEE Netw.}, vol.~33, no.~4, pp. 70--75,
  Jul./Aug. 2019.

\bibitem{6G3}
Z.~Yang, Y.~Li, Y.~L. Guan, and Y.~Fang, ``Source-constrained hierarchical
  modulation systems with protograph {LDPC} codes: A promising transceiver
  design for future 6{G}-enabled {I}o{T},'' \emph{IEEE J. Sel. Areas Commun.},
  vol.~43, no.~4, pp. 1103--1117, Apr. 2025.

\bibitem{reconfigure1}
S.~Gong, X.~Lu, D.~T. Hoang, D.~Niyato, L.~Shu, D.~I. Kim, and Y.-C. Liang,
  ``Toward smart wireless communications via intelligent reflecting surfaces: A
  contemporary survey,'' \emph{IEEE Commun. Surveys \& Tuts.}, vol.~22, no.~4,
  pp. 2283--2314, 4th Quart. 2020.

\bibitem{reconfigure2}
B.~A. Cetiner, E.~Akay, E.~Sengul, and E.~Ayanoglu, ``A {MIMO} system with
  multifunctional reconfigurable antennas,'' \emph{IEEE Antennas Wireless
  Propag. Lett.}, vol.~5, pp. 463--466, s 2006.

\bibitem{reconfigure3}
A.~M. Sayeed and V.~Raghavan, ``Maximizing {MIMO} capacity in sparse multipath
  with reconfigurable antenna arrays,'' \emph{IEEE J. Sel. Topics Signal
  Proc.}, vol.~1, no.~1, pp. 156--166, Jun. 2007.

\bibitem{reconfigure4}
J.~Aberle, S.-H. Oh, D.~Auckland, and S.~Rogers, ``Reconfigurable antennas for
  wireless devices,'' \emph{IEEE Antenna Propag. Mag.,}, vol.~45, no.~6, pp.
  148--154, Dec. 2003.

\bibitem{AS1}
S.~Sanayei and A.~Nosratinia, ``Antenna selection in {MIMO} systems,''
  \emph{IEEE Commun. Mag.}, vol.~42, no.~10, pp. 68--73, Oct. 2004.

\bibitem{AS2}
A.~Molisch and M.~Win, ``{MIMO} systems with antenna selection,'' \emph{IEEE
  Microw. Mag.}, vol.~5, no.~1, pp. 46--56, Mar. 2004.

\bibitem{AS3}
Y.~Gao, H.~Vinck, and T.~Kaiser, ``Massive {MIMO} antenna selection: Switching
  architectures, capacity bounds, and optimal antenna selection algorithms,''
  \emph{IEEE Trans. Signal Process.}, vol.~66, no.~5, pp. 1346--1360, Mar.
  2018.

\bibitem{AS4}
S.~Asaad, A.~M. Rabiei, and R.~R. Müller, ``Massive {MIMO} with antenna
  selection: Fundamental limits and applications,'' \emph{IEEE Trans. Wireless
  Commun.}, vol.~17, no.~12, pp. 8502--8516, Dec. 2018.

\bibitem{RIS1}
S.~Gong, X.~Lu, D.~T. Hoang, D.~Niyato, L.~Shu, D.~I. Kim, and Y.-C. Liang,
  ``Toward smart wireless communications via intelligent reflecting surfaces: A
  contemporary survey,'' \emph{IEEE Commun. Surveys \& Tut.}, vol.~22, no.~4,
  pp. 2283--2314, 4th Quart. 2020.

\bibitem{RIS2}
E.~Basar, M.~Di~Renzo, J.~De~Rosny, M.~Debbah, M.-S. Alouini, and R.~Zhang,
  ``Wireless communications through reconfigurable intelligent surfaces,''
  \emph{IEEE Access}, vol.~7, pp. 116\,753--116\,773, 2019.

\bibitem{FA1}
K.-K. Wong, A.~Shojaeifard, K.-F. Tong, and Y.~Zhang, ``Fluid antenna
  systems,'' \emph{IEEE Trans. Wireless Commun.}, vol.~20, no.~3, pp.
  1950--1962, Mar. 2021.

\bibitem{MA}
L.~Zhu, W.~Ma, and R.~Zhang, ``Modeling and performance analysis for movable
  antenna enabled wireless communications,'' \emph{IEEE Trans. Wireless
  Commun.}, vol.~23, no.~6, pp. 6234--6250, Jun. 2024.

\bibitem{PA1}
Z.~Ding, R.~Schober, and H.~Vincent~Poor, ``Flexible-antenna systems: A
  pinching-antenna perspective,'' \emph{IEEE Trans. on Commun.}, pp. 1--1,
  2025, early access.

\bibitem{AS_new1}
A.~Molisch, M.~Win, Y.-S. Choi, and J.~Winters, ``Capacity of {MIMO} systems
  with antenna selection,'' \emph{IEEE Trans. Wireless Commun.}, vol.~4, no.~4,
  pp. 1759--1772, Jul. 2005.

\bibitem{AS_new2}
S.~Cui, A.~Goldsmith, and A.~Bahai, ``Energy-efficiency of {MIMO} and
  cooperative {MIMO} techniques in sensor networks,'' \emph{IEEE J. Sel. Areas
  Commun.}, vol.~22, no.~6, pp. 1089--1098, Aug. 2004.

\bibitem{AS_new3}
X.~Zhou, B.~Bai, and W.~Chen, ``Iterative antenna selection for multi-stream
  {MIMO} under a holistic power model,'' \emph{IEEE Wireless Commun. Lett.},
  vol.~3, no.~1, pp. 82--85, Feb. 2014.

\bibitem{AS_new4}
J.~Xu and L.~Qiu, ``Energy efficiency optimization for {MIMO} broadcast
  channels,'' \emph{IEEE Trans. Wireless Commun.}, vol.~12, no.~2, pp.
  690--701, Feb. 2013.

\bibitem{AS_new5}
H.~Li, L.~Song, D.~Zhu, and M.~Lei, ``Energy efficiency of large scale mimo
  systems with transmit antenna selection,'' in \emph{Proc. IEEE Int. Conf.
  Commun.}, Budapest, Hungary, Jun. 2013, pp. 4641--4645.

\bibitem{RIS_new1}
L.~Subrt and P.~Pechac, ``Intelligent walls as autonomous parts of smart indoor
  environments,'' \emph{IET Commun.,}, vol.~6, pp. 1004--1010, May 2012.

\bibitem{RIS_new2}
Q.~Wu and R.~Zhang, ``Intelligent reflecting surface enhanced wireless network:
  Joint active and passive beamforming design,'' in \emph{Proc. IEEE Global
  Commun. Conf. (GLOBECOM)}, Abu Dhabi, United Arab Emirates, Dec. 2018, pp.
  1--6.

\bibitem{RIS_new3}
{\"O}.~{\"O}zdogan, E.~Bj{\"o}rnson, and E.~G. Larsson, ``Using intelligent
  reflecting surfaces for rank improvement in {MIMO} communications,'' in
  \emph{proc. IEEE ICASSP}, Barcelona, Spain, May 2020, pp. 9160--9164.

\bibitem{RIS_new4}
Y.~Liu, L.~Zhang, B.~Yang, W.~Guo, and M.~A. Imran, ``Programmable wireless
  channel for multi-user {MIMO} transmission using meta-surface,'' in
  \emph{Proc. IEEE Global Commun. Conf. (GLOBECOM)}, Waikoloa, HI, USA, Dec.
  2019, pp. 1--6.

\bibitem{RIS_new5}
Q.~Wu and R.~Zhang, ``Intelligent reflecting surface enhanced wireless network
  via joint active and passive beamforming,'' \emph{IEEE Trans. Wireless
  Commun.}, vol.~18, no.~11, pp. 5394--5409, Nov. 2019.

\bibitem{RIS_new6}
X.~Yu, D.~Xu, and R.~Schober, ``Enabling secure wireless communications via
  intelligent reflecting surfaces,'' in \emph{Proc. IEEE Global Commun. Conf.
  (GLOBECOM)}, Waikoloa, HI, USA, Dec. 2019, pp. 1--6.

\bibitem{FA_new1}
W.~K. New, K.-K. Wong, H.~Xu, K.-F. Tong, and C.-B. Chae, ``An
  information-theoretic characterization of {MIMO-FAS}: Optimization,
  diversity-multiplexing tradeoff and q-outage capacity,'' \emph{IEEE Trans.
  Wireless Commun.}, vol.~23, no.~6, pp. 5541--5556, Jun. 2024.

\bibitem{FA_new2}
K.-K. Wong and K.-F. Tong, ``Fluid antenna multiple access,'' \emph{IEEE Trans.
  Wireless Commun.}, vol.~21, no.~7, pp. 4801--4815, Jul. 2022.

\bibitem{FA_new3}
K.-K. Wong, D.~Morales-Jimenez, K.-F. Tong, and C.-B. Chae, ``Slow fluid
  antenna multiple access,'' \emph{IEEE Trans. Commun.}, vol.~71, no.~5, pp.
  2831--2846, May 2023.

\bibitem{FA_new4}
J.~Zheng, J.~Zhang, H.~Du, D.~Niyato, S.~Sun, B.~Ai, and K.~B. Letaief,
  ``Flexible-position {MIMO} for wireless communications: Fundamentals,
  challenges, and future directions,'' \emph{IEEE Wireless Commun.}, vol.~31,
  no.~5, pp. 18--26, Oct. 2024.

\bibitem{FA_new5}
L.~Zhu, W.~Ma, and R.~Zhang, ``Movable-antenna array enhanced beamforming:
  Achieving full array gain with null steering,'' \emph{IEEE Commun. Lett.},
  vol.~27, no.~12, pp. 3340--3344, Dec. 2023.

\bibitem{PA_new1}
H.~O.~Y. Suzuki and K.~Kawai, ``Pinching antenna: Using a dielectric waveguide
  as an antenna,'' \emph{NTT DOCOMO Technical J.}, vol.~23, no.~4, pp. 5--12,
  Jan. 2022.

\bibitem{PA_new2}
C.~Ouyang, Z.~Wang, Y.~Liu, and Z.~Ding, ``Array gain for pinching-antenna
  systems ({PASS}),'' \emph{IEEE Commun. Lett.}, pp. 1--1, 2025, early access.

\bibitem{PA_new3}
Y.~Xu, Z.~Ding, and G.~K. Karagiannidis, ``Rate maximization for downlink
  pinching-antenna systems,'' \emph{IEEE Wireless Commun. Lett.}, vol.~14,
  no.~5, pp. 1431--1435, May 2025.

\bibitem{PA_new4}
K.~Wang, Z.~Ding, and R.~Schober, ``Antenna activation for {NOMA} assisted
  pinching-antenna systems,'' \emph{IEEE Wireless Commun. Lett.}, vol.~14,
  no.~5, pp. 1526--1530, May 2025.

\bibitem{PA_new5}
S.~A. Tegos, P.~D. Diamantoulakis, Z.~Ding, and G.~K. Karagiannidis, ``Minimum
  data rate maximization for uplink pinching-antenna systems,'' \emph{IEEE
  Wireless Commun. Lett.}, vol.~14, no.~5, pp. 1516--1520, May 2025.

\bibitem{PA_new6}
J.-C. Chen, P.-C. Wu, and K.-K. Wong, ``Dynamic placement of pinching antennas
  for multicast {MU-MISO} downlinks,'' \emph{IEEE Open J. Commun. Soc.}, pp.
  1--1, 2025, early access.

\bibitem{IM1}
E.~Ba\c{s}ar, U.~Ayg\"ol\"u, E.~Panay{\i}rc{\i}, and H.~V. Poor, ``Orthogonal
  frequency division multiplexing with index modulation,'' \emph{IEEE Trans.
  Signal Process.}, vol.~61, no.~22, pp. 5536--5549, Nov. 2013.

\bibitem{IM2}
A.~A. Purwita, A.~Yesilkaya, T.~Cogalan, M.~Safari, and H.~Haas, ``Generalized
  time slot index modulation for {LiFi},'' in \emph{Proc. IEEE 30th Annu. Int.
  Symp. Pers., Indoor Mobile Radio Commun. (PIMRC)}, Istanbul, Turkey, Sep.
  2019, pp. 1--7.

\bibitem{IM3}
{R. Y. Mesleh}, {H. Haas}, {S. Sinanovic}, {C. W. Ahn}, and {S. Yun}, ``Spatial
  modulation,'' \emph{{IEEE} Trans. Veh. Technol.}, vol.~57, no.~4, pp.
  2228--2241, Jul. 2008.

\bibitem{IM4}
M.~Di~Renzo, H.~Haas, A.~Ghrayeb, S.~Sugiura, and L.~Hanzo, ``Spatial
  modulation for generalized {MIMO}: Challenges, opportunities, and
  implementation,'' \emph{Proc. IEEE}, vol. 102, no.~1, pp. 56--103, Jan. 2014.

\bibitem{IM5}
P.~Yang, M.~Di~Renzo, Y.~Xiao, S.~Li, and L.~Hanzo, ``Design guidelines for
  spatial modulation,'' \emph{IEEE Commun. Surveys Tuts.}, vol.~17, no.~1, pp.
  6--26, 1st Quart. 2015.

\bibitem{SM1}
E.~Ba{\c s}ar, U.~Ayg{\"o}l{\"u}, E.~Panayirci, and H.~V. Poor, ``Space-time
  block coded spatial modulation,'' \emph{{IEEE} Trans. Wireless Commun.},
  vol.~59, no.~3, pp. 823--832, Mar. 2011.

\bibitem{SM2}
{L. Xiao}, {Y. Xiao}, {C. Xu}, {X. Lei}, {P. Yang}, {S. Li}, and {L. Hanzo},
  ``Compressed-sensing assisted spatial multiplexing aided spatial
  modulation,'' \emph{{IEEE} Trans. Wireless Commun.}, vol.~17, no.~2, pp.
  794--807, Feb. 2018.

\bibitem{SM3}
P.~Wolniansky, G.~Foschini, G.~Golden, and R.~Valenzuela, ``V-{{BLAST}}: An
  architecture for realizing very high data rates over the rich-scattering
  wireless channel,'' in \emph{Proc. URSI ISSSE}, Pisa, Italy, Sep. 1998, pp.
  295--300.

\bibitem{SM4}
E.~Basar, ``Reconfigurable intelligent surface-based index modulation: A new
  beyond mimo paradigm for {6G},'' \emph{IEEE Trans. Commun.}, vol.~68, no.~5,
  pp. 3187--3196, May. 2020.

\bibitem{BO}
X.-w. Chang and Q.~Han, ``Solving box-constrained integer least squares
  problems,'' \emph{IEEE Trans. Wireless Commun.}, vol.~7, no.~1, pp. 277--287,
  Jan. 2008.

\bibitem{active_set}
W.~W. Hager and H.~Zhang, ``A new active set algorithm for box constrained
  optimization,'' \emph{SIAM J. Optim.}, vol.~17, no.~2, pp. 526--557, 2006.

\bibitem{complexity}
U.~Fincke and M.~Pohst, ``Improved methods for calculating vectors of short
  length in a lattice, including a complexity analysis,'' \emph{Mathematics of
  Computation}, vol.~44, no. 170, pp. 463--471, Apr. 1985.

\bibitem{upperbound}
M.~K. Simon and M.-S. Alouini, \emph{Digital Communication over Fading
  Channels}, 1st~ed.\hskip 1em plus 0.5em minus 0.4em\relax New York: Wiley,
  2001.

\bibitem{Qfunction}
------, \emph{Digital Communication over Fading Channels: A Unified Approach to
  Performance Analysis}.\hskip 1em plus 0.5em minus 0.4em\relax {New York}:
  {John Wiley \& Sons}, 2000.

\bibitem{MGF}
G.~L. Turin, ``The characteristic function of hermitian quadratic forms in
  complex normal variables,'' \emph{Biometrika}, vol.~47, no. 1-2, pp.
  199--201, Jun. 1960.

\bibitem{channel_estimation1}
\BIBentryALTinterwordspacing
G.~Zhou, V.~Papanikolaou, Z.~Ding, and R.~Schober, ``Channel estimation for
  mm{W}ave pinching-antenna systems,'' 2025. [Online]. Available:
  \url{https://arxiv.org/abs/2504.09317}
\BIBentrySTDinterwordspacing

\bibitem{channel_estimation2}
J.~Xiao, J.~Wang, and Y.~Liu, ``Channel estimation for pinching-antenna systems
  ({PASS}),'' \emph{IEEE Commun. Lett.}, pp. 1--1, 2025, early access.

\bibitem{Armijo}
D.~P. Bertsekas, ``Nonlinear programming,'' \emph{J. Oper. Res. Soc.}, vol.~48,
  no.~3, pp. 334--334, Dec. 1997.

\bibitem{convergence}
P.-A. Absil, R.~Mahony, and R.~Sepulchre, ``Optimization algorithms on matrix
  manifolds,'' in \emph{Optimization Algorithms on Matrix Manifolds}.\hskip 1em
  plus 0.5em minus 0.4em\relax Princeton University Press, 2009.

\bibitem{3gpp}
\textit{3rd Generation Partnership Project; Technical Specification Group Radio
  Access Network}; \textit{Spatial Channel Model for Multiple Input Multiple
  Output (MIMO) Simulations}, 3GPP TR 25.996 V14.0.0, Mar. 2017.

\bibitem{LoS1}
O.~\"Ozdogan, E.~Bj\"ornson, and J.~Zhang, ``Cell-free massive {MIMO} with
  {Rician} fading: Estimation schemes and spectral efficiency,'' in \emph{Proc.
  52nd Asilomar Conf. Signals, Syst., Comput.}, Pacific Grove, CA, USA, Oct.
  2018, pp. 975--979.

\bibitem{LoS2}
O.~\"Ozdogan, E.~Bj\"ornson, and E.~G. Larsson, ``Massive {MIMO} with spatially
  correlated {Rician} fading channels,'' \emph{IEEE Trans. Commun.}, vol.~67,
  no.~5, pp. 3234--3250, May 2019.

\end{thebibliography}

\vfill

\end{document}